\documentclass[12pt,draftclsnofoot,journal,onecolumn]{IEEEtran}

\usepackage{cite}
\usepackage{color}
\usepackage{threeparttable}
\usepackage{graphicx}
\usepackage{picinpar}
\usepackage{epstopdf}
\usepackage[cmex10]{amsmath}
\usepackage{amsmath,amsfonts,amssymb}
\usepackage{subfigure}
\usepackage{changepage}
\usepackage{algorithm}
\usepackage{caption} 
\usepackage{algpseudocode}
\usepackage{stfloats}
\usepackage{bm}
\usepackage{amsthm}
\usepackage{enumerate}
\usepackage{multirow}
\usepackage{booktabs}
\usepackage{url}
\usepackage{diagbox}

\usepackage{hyperref}
\usepackage{pifont}

\allowdisplaybreaks



\begin{document}
\newtheorem{remark}{Remark}
\newtheorem{lemma}{Lemma}
\newtheorem{corollary}{Corollary}
\newtheorem{theorem}{Theorem}
\newtheorem{proposition}{Proposition}
\newtheorem{definition}{Definition}
\newcommand{\e}{\begin{equation}}
\newcommand{\ee}{\end{equation}}
\newcommand{\eqn}{\begin{eqnarray}}
\newcommand{\eeqn}{\end{eqnarray}}
\newenvironment{shrinkeq}[1]
{ \bgroup
\addtolength\abovedisplayshortskip{#1}
\addtolength\abovedisplayskip{#1}
\addtolength\belowdisplayshortskip{#1}
\addtolength\belowdisplayskip{#1}}
{\egroup\ignorespacesafterend}
\title{\LARGE Integrated Sensing and Communication with mmWave Massive MIMO: A Compressed Sampling Perspective}

\author{Zhen Gao, Ziwei Wan, Dezhi Zheng, Shufeng Tan, Christos Masouros, Derrick Wing Kwan Ng, \IEEEmembership{Fellow,~IEEE}, and Sheng Chen, \IEEEmembership{Fellow,~IEEE}

}

\maketitle

\vspace{-10mm}

\begin{abstract}
Integrated sensing and communication (ISAC) has opened up numerous game-changing opportunities for realizing future wireless systems. In this paper, we propose an ISAC processing framework relying on millimeter-wave (mmWave) massive multiple-input multiple-output (MIMO) systems. Specifically, we provide a compressed sampling (CS) perspective to facilitate ISAC processing, which can not only recover the high-dimensional channel state information or/and radar imaging information, but also significantly reduce pilot overhead. First, an energy-efficient widely spaced array (WSA) architecture is tailored for the radar receiver, which enhances the angular resolution of radar sensing at the cost of angular ambiguity. Then, we propose an ISAC frame structure for time-varying ISAC systems considering different timescales. The pilot waveforms are judiciously designed by taking into account both CS theories and hardware constraints induced by hybrid beamforming (HBF) architecture. Next, we design the dedicated dictionary for WSA that serves as a building block for formulating the ISAC processing as sparse signal recovery problems. The orthogonal matching pursuit with support refinement (OMP-SR) algorithm is proposed to effectively solve the problems in the existence of the angular ambiguity. We also provide a framework for estimating the Doppler frequencies during payload data transmission to guarantee communication performances. Simulation results demonstrate the good performances of both communications and radar sensing under the proposed ISAC framework.
\end{abstract}


\IEEEpeerreviewmaketitle

\section{Introduction}\label{S1}

As two representative applications of radio technology, {\it wireless communications} and {\it radar sensing} have respectively achieved remarkable results over the past few decades. Given the spectrum crunch caused by ever-increasing connected devices and applications, integrated sensing and communication (ISAC) has attracted great research interest recently \cite{CM1,CM2,A,ALiu,C,TcomSurvey}. On the one hand, ISAC allows communication systems and radar systems to share the scarce spectrum and expensive hardware resources, saving a large amount of cost. On the other hand, some critical scenarios in beyond fifth-generation (B5G) and even 6G, such as autonomous driving \cite{MagaVT}, Wi-Fi sensing \cite{Wi-Fi}, and extended reality \cite{Nature}, require resilient communications together with high-precision environment sensing ability provided by advanced radar techniques. Therefore, ISAC is expected to benefit both communication and radar communities in the near future.
	
In this paper, we focus on combining radar systems with state-of-the-art wireless communications. Specifically, millimeter-wave (mmWave) massive multiple-input multiple-output (mMIMO) system, which is the backbone of physical-layer techniques in 5G and beyond, is also an attractive solution for radar sensing. Capitalizing on the high angular resolution of mmWave mMIMO, the shapes of the targets can be clearly identified by exploiting the radar echo signals from different directions. Hence, target imaging (rather than simply detecting the existence of a target) via radio frequency (RF) signals can be realized \cite{AccessImaging}.
As such, we aim to integrate radar sensing into the mmWave channel estimation (CE).

Despite the ambitious visions, a critical issue in mMIMO-aided ISAC systems is that a massive number of antennas introduce a significant computational burden on the signal processing for both communications and radar sensing. This issue becomes more pronounced when target imaging is considered, since a radar system needs to recover the high-dimensional imaging information instead of estimating a few parameters of the targets. Besides, low pilot overhead is a prerequisite for effective time-varying CE so that high-mobility users in ISAC scenarios can be reliably served. Given these challenges, compressed sampling (CS), also known as compressive sensing, which can recover signals from reduced measurements by leveraging the intrinsic sparsity \cite{GaoCM}, is a promising solution for ISAC systems. Although CS techniques have already spread rapidly in many disciplines, it is necessary to further study their applications to ISAC systems.

\subsection{Prior Work}\label{S1.1}

Radar systems assisted by antenna arrays are usually divided into two basic types, {\it phased-array radar} and {\it MIMO radar} \cite{TwoTypes}, based on whether the transmit waveforms at different antennas are coherent or not. To combine the advantages of these two types of radars, the authors of \cite{Tradeoff} proposed the concept of {\it phased-MIMO radar}, which can achieve both the waveform diversity and coherent processing gain. Moreover, sophisticated antenna array forms, such as virtual uniform linear array (ULA) \cite{STAP}, nested array \cite{Nest}, or co-prime array \cite{Co-prime}, have been proposed and deployed at radar transceivers for enhancing the degrees of freedoms. As an example, the virtual ULA \cite{STAP} considered a radar array with $N_{\rm T}$ transmit antennas and $N_{\rm R}$ receive antennas, both uniformly spaced. It was shown that by widening the antenna spacing of transmit array, a virtual array with $N_{\rm T}N_{\rm R}$ effective aperture can be obtained with $N_{\rm T}+N_{\rm R}$ antennas. This architecture has been realized in practical radar chips for commercial use (cf. \cite[Fig. 2]{AccessImaging}). However, this widely-spaced transmit array is unsuitable when communication functionality is needed, since it will cause the spatial aliasing due to sub-Nyquist spatial sampling \cite{CSradar3}, introducing the interferences for beamforming design. 
Links between radars and CS have been explored in \cite{CSradar1,CSradar2,CSradar3,CSradar4,CSradar5,CSradar6}. The motivation behind CS approaches is that the intrinsic ``sparse'' nature exists in many radar problems (e.g., the limited number of interested targets, or the compressibility of radar images). On that basis, the CS algorithms have been applied to the parameter estimation under the single-input single-output radar \cite{CSradar1}, the MIMO radar with Nyquist spatial sampling array \cite{CSradar2}, or the MIMO radar with sub-Nyquist spatial sampling array \cite{CSradar3,CSradar4}, and to the radar image reconstruction under synthetic aperture radar (SAR) systems \cite{CSradar5,CSradar6}.

In the theory and practice of communications, mmWave mMIMO with {\it hybrid beamforming} (HBF) architecture \cite{Hybrid} has been considered as a key enabler for 5G/B5G. By connecting large-scale array with a few RF chains (RFCs) through a fully or partially connected phase shifter network, the HBF architecture realizes the trade-off between the hardware complexity and the system performance. Nevertheless, this hybrid architecture significantly decreases the dimension of the received signals, imposing great challenges to CE for mmWave mMIMO. As a remedy, a hardware solution was introduced in \cite{MagaADC}, where a fully-digital receiver with low-resolution analog-to-digital converters (ADCs) is employed to reserve the high-dimensional received signals at the cost of severe quantization noises. Moreover, CS techniques have also been widely adopted for wireless communications.
Based on the sparsity of mmWave channels in the angular domain \cite{GaoCL,korean,MyTVT}, the delay domain \cite{XMa}, or a mixture of both \cite{HeathJSAC,XLin}, the literature \cite{MyTVT,GaoCL,korean,XMa,HeathJSAC,XLin} formulated the mmWave CE as the corresponding sparse signal recovery problems, using off-the-shelf CS-based algorithms. In addition to CE, beamforming design can also be well supported by CS techniques. The authors of \cite{SpatialSparse} proposed a beamforming scheme for mmWave mMIMO with HBF architecture. In particular, the CS-based algorithm was exploited to reconstruct both the RF precoder and the baseband precoder so that their combination can mimic the optimal fully-digital precoder. Since then, the idea of \cite{SpatialSparse} has further extended to more communication scenarios, e.g.,  \cite{Adaptive,KeKe}. However, these CS-based methods are dedicated only for communications, and they may not be straightforwardly applicable in ISAC scenarios.

For ISAC applications, research efforts towards dual-functional radar-communication (DFRC) are well underway. In the literature {\cite{CM3,FLiuTWC,SARJRC,AZhang,CSJRC1,CSJRC2,Similar}}, DFRC integrates radar and communication signals in the temporal, the frequency, or the spatial domain, which is deemed more beneficial than its single constituent part (communications or radar).
A demonstration of airborne MIMO radar was presented in \cite{SARJRC}, and the field test was conducted to validate the feasibility of joint communication and SAR imaging.
In \cite{FLiuTWC}, the authors investigated a radar-assisted predictive beamforming design for vehicular networks. The information obtained by radar sensing is fed to the design of communication beam tacking, to improve the performance.
Also, the authors of \cite{Similar} optimized the DFRC transmit sequences with one-bit digital-to-analog converters (DACs) to ensure the performances of both symbol demodulation and radar detection. Unfortunately, the works \cite{SARJRC,FLiuTWC,Similar} do not consider the advanced HBF architecture, which has been regarded as a promising candidate for ISAC. In fact, the concept of HBF is quite similar to that of phased-MIMO radar. Inspired by this observation, a comprehensive ISAC framework based on HBF architecture was initially proposed in \cite{TcomSurvey}. An orthogonal waveform was applied in \cite{TcomSurvey} to minimize the Cram{\'e}r-Rao bound (CRB) of parameter estimation. Yet, it imposes unaffordable pilot overhead and ignores the practical hardware constraint of HBF architecture. 
The application of CS to ISAC can be found in \cite{AZhang,CSJRC1,CSJRC2}, which provides insight in reducing the pilot overhead for ISAC processing. In \cite{AZhang}, an ISAC framework relying on analog beamforming was proposed. With the designed multibeam for both communications and radar sensing, the authors of \cite{AZhang} adopted multiple measurement vector (MMV) CS to estimate the parameters of radar targets.
Later, the authors of \cite{CSJRC1} studied radar sensing with one-dimension (1D) to 3D CS techniques, using the signals compatible with 5G standards. This work \cite{CSJRC1} has been extended in \cite{CSJRC2}, where a background subtraction method was proposed to reduce the clutter in the input signals, benefiting CS algorithms further.
However, \cite{AZhang,CSJRC2} were based on the impractical on-grid parameter model, and \cite{AZhang,CSJRC1,CSJRC2} did not consider the mmWave mMIMO with HBF architecture and the corresponding waveform design.

\begin{table*}[!tp]
\captionsetup{font={footnotesize,color={black}},labelsep=newline}
\caption{\scshape The Summary of Literature Review}
\vspace{-2mm}
\centering
\tiny

\begin{tabular}{|l|cccccl|cccc|}
\hline
\multicolumn{1}{|c|}{\multirow{2}{*}{\diagbox{\bf Features}{\bf Works}}} &
  \multicolumn{6}{c|}{\bf Radar-Only} &
  \multicolumn{4}{c|}{\bf Commnication-Only} \\ \cline{2-11} 
\multicolumn{1}{|c|}{} &
  \multicolumn{1}{c|}{\cite{AccessImaging}} &
  \multicolumn{1}{c|}{\cite{STAP,Nest,Co-prime}} &
  \multicolumn{1}{c|}{\cite{CSradar1}} &
  \multicolumn{1}{c|}{\cite{CSradar2}} &
  \multicolumn{1}{c|}{\cite{CSradar3,CSradar4}} &
  \multicolumn{1}{c|}{\cite{CSradar5,CSradar6}} &
  \multicolumn{1}{c|}{\cite{MagaADC,GaoCL,HeathJSAC,XLin,KeKe,SpatialSparse}} &
  \multicolumn{1}{c|}{\cite{korean,Adaptive}} &
  \multicolumn{1}{c|}{\cite{MyTVT}} &
  {\cite{XMa}} \\ \hline
\bf Designed Array &
  \multicolumn{1}{c|}{\checkmark} &
  \multicolumn{1}{c|}{\checkmark} &
  \multicolumn{1}{c|}{} &
  \multicolumn{1}{c|}{\checkmark} &
  \multicolumn{1}{c|}{\checkmark} &
   &
  \multicolumn{1}{c|}{} &
  \multicolumn{1}{c|}{} &
  \multicolumn{1}{c|}{\checkmark} &
   \\ \hline
\bf CS Techniques &
  \multicolumn{1}{c|}{} &
  \multicolumn{1}{c|}{} &
  \multicolumn{1}{c|}{\checkmark} &
  \multicolumn{1}{c|}{\checkmark} &
  \multicolumn{1}{c|}{\checkmark} &
  \multicolumn{1}{c|}{\checkmark} &
  \multicolumn{1}{c|}{\checkmark} &
  \multicolumn{1}{c|}{\checkmark} &
  \multicolumn{1}{c|}{\checkmark} &
  \checkmark \\ \hline
\bf 3D$^1$ Sensing &
  \multicolumn{1}{c|}{} &
  \multicolumn{1}{c|}{} &
  \multicolumn{1}{c|}{} &
  \multicolumn{1}{c|}{\checkmark} &
  \multicolumn{1}{c|}{} &
   &
  \multicolumn{1}{c|}{} &
  \multicolumn{1}{c|}{} &
  \multicolumn{1}{c|}{} &
   \\ \hline
\bf Target Imaging &
  \multicolumn{1}{c|}{\checkmark} &
  \multicolumn{1}{c|}{} &
  \multicolumn{1}{c|}{} &
  \multicolumn{1}{c|}{} &
  \multicolumn{1}{c|}{} &
  \multicolumn{1}{c|}{\checkmark} &
  \multicolumn{1}{c|}{} &
  \multicolumn{1}{c|}{} &
  \multicolumn{1}{c|}{} &
   \\ \hline
\bf mmWave mMIMO &
  \multicolumn{1}{c|}{\checkmark} &
  \multicolumn{1}{c|}{} &
  \multicolumn{1}{c|}{} &
  \multicolumn{1}{c|}{} &
  \multicolumn{1}{c|}{} &
   &
  \multicolumn{1}{c|}{\checkmark} &
  \multicolumn{1}{c|}{\checkmark} &
  \multicolumn{1}{c|}{\checkmark} &
  \checkmark \\ \hline
\bf HBF Architecture &
  \multicolumn{1}{c|}{} &
  \multicolumn{1}{c|}{} &
  \multicolumn{1}{c|}{} &
  \multicolumn{1}{c|}{} &
  \multicolumn{1}{c|}{} &
   &
  \multicolumn{1}{c|}{\checkmark} &
  \multicolumn{1}{c|}{\checkmark} &
  \multicolumn{1}{c|}{\checkmark} &
   \\ \hline
\bf Waveform Design$^2$ &
  \multicolumn{1}{c|}{} &
  \multicolumn{1}{c|}{} &
  \multicolumn{1}{c|}{} &
  \multicolumn{1}{c|}{} &
  \multicolumn{1}{c|}{} &
   &
  \multicolumn{1}{c|}{} &
  \multicolumn{1}{c|}{\checkmark} &
  \multicolumn{1}{c|}{\checkmark} &
  \checkmark \\ \hline
\bf Frame Structure Design &
  \multicolumn{1}{c|}{} &
  \multicolumn{1}{c|}{} &
  \multicolumn{1}{c|}{} &
  \multicolumn{1}{c|}{} &
  \multicolumn{1}{c|}{} &
   &
  \multicolumn{1}{c|}{} &
  \multicolumn{1}{c|}{} &
  \multicolumn{1}{c|}{} &
   \\ \hline
\end{tabular}

\vspace{1mm}

\begin{tabular}{|l|cccccc|c|}
\hline
\multicolumn{1}{|c|}{\multirow{2}{*}{\diagbox{\bf Features}{\bf Works}}} &
  \multicolumn{6}{c|}{\bf ISAC} &
  \multirow{2}{*}{\bf Our Work} \\ \cline{2-7}
\multicolumn{1}{|c|}{} &
  \multicolumn{1}{c|}{\cite{TcomSurvey}} &
  \multicolumn{1}{c|}{\cite{CM3,Similar}} &
  \multicolumn{1}{c|}{\cite{SARJRC}} &
  \multicolumn{1}{c|}{\cite{FLiuTWC}} &
  \multicolumn{1}{c|}{\cite{AZhang}} &
  {\cite{CSJRC1,CSJRC2}} &
   \\ \hline
\bf Designed Array &
  \multicolumn{1}{l|}{} &
  \multicolumn{1}{c|}{} &
  \multicolumn{1}{c|}{} &
  \multicolumn{1}{c|}{} &
  \multicolumn{1}{c|}{} &
   &
  \checkmark \\ \hline
\bf CS Techniques &
  \multicolumn{1}{l|}{} &
  \multicolumn{1}{c|}{} &
  \multicolumn{1}{c|}{} &
  \multicolumn{1}{c|}{} &
  \multicolumn{1}{c|}{\checkmark} &
  \checkmark &
  \checkmark \\ \hline
\bf 3D$^1$ Sensing &
  \multicolumn{1}{c|}{\checkmark} &
  \multicolumn{1}{c|}{} &
  \multicolumn{1}{c|}{} &
  \multicolumn{1}{c|}{} &
  \multicolumn{1}{c|}{\checkmark} &
  \checkmark &
  \checkmark \\ \hline
\bf Target Imaging &
  \multicolumn{1}{l|}{} &
  \multicolumn{1}{c|}{} &
  \multicolumn{1}{c|}{\checkmark} &
  \multicolumn{1}{c|}{} &
  \multicolumn{1}{c|}{} &
   &
  \checkmark \\ \hline
\bf mmWave mMIMO &
  \multicolumn{1}{c|}{\checkmark} &
  \multicolumn{1}{c|}{} &
  \multicolumn{1}{c|}{} &
  \multicolumn{1}{c|}{\checkmark} &
  \multicolumn{1}{c|}{} &
   &
  \checkmark \\ \hline
\bf HBF Architecture &
  \multicolumn{1}{c|}{\checkmark} &
  \multicolumn{1}{c|}{} &
  \multicolumn{1}{c|}{} &
  \multicolumn{1}{c|}{} &
  \multicolumn{1}{c|}{} &
   &
  \checkmark \\ \hline
\bf Waveform Design$^2$ &
  \multicolumn{1}{c|}{\checkmark} &
  \multicolumn{1}{c|}{\checkmark} &
  \multicolumn{1}{c|}{\checkmark} &
  \multicolumn{1}{c|}{\checkmark} &
  \multicolumn{1}{c|}{\checkmark} &
   &
  \checkmark \\ \hline
\bf Frame Structure Design &
  \multicolumn{1}{l|}{} &
  \multicolumn{1}{c|}{} &
  \multicolumn{1}{c|}{} &
  \multicolumn{1}{c|}{\checkmark} &
  \multicolumn{1}{c|}{\checkmark} &
   &
  \checkmark \\ \hline
\end{tabular}

\begin{tablenotes}
	\scriptsize
	\item $^1$ ``3D'' refers to ``Angle-Range-Doppler''.
	\ \ \ \ $^2$ This includes the pilot design for CE.
\end{tablenotes}
\vspace{-10mm}
\end{table*}

\subsection{Our Contributions}\label{S1.2}

Based on the aforementioned discussions, it is evident that the applications of mmWave mMIMO and CS techniques to ISAC scenarios are still at an early development stage. In this work, we focus on the ISAC systems relying on mmWave mMIMO and the application of advanced CS techniques. 
A brief summary of the existing and our proposed works is presented in Table I.
To be specific, our main contributions can be summarized as follows. 

\begin{itemize}
\item {\bf We propose a DFRC transceiver architecture based on mmWave mMIMO.} 
The proposed architecture consists of one communication unit (CU) and one radar unit (RU). A critically spaced array (CSA) with the HBF architecture is adopted at the CU, while a widely spaced array (WSA) with low-resolution ADCs is designed for the RU to receive the radar echo signals. The proposed architecture guarantees both the energy-efficient communications and the high angular resolution for radar sensing, at the cost of the angular ambiguity.
\item {\bf We study the integration of radar sensing into the CE of the conventional cellular communications by proposing an ISAC frame structure and a waveform design scheme tailored for HBF architecture and CS processing.} By considering different timescales, the proposed frame structure can cope with fast time-varying environments, where high-mobility targets can be tracked and high-mobility users can be served. The proposed pilot waveform design not only sufficiently diversifies the pilot waveform, as required by CS theory, but also satisfies the hardware constraints imposed by the HBF architecture.
\item {\bf We propose a dedicated CS-based algorithm to overcome the angular ambiguity brought by the WSA.} By leveraging the natural spatial consistence, i.e., the co-located CU and RU observe targets at the same directions, we propose the orthogonal matching pursuit with support refinement (OMP-SR) algorithm for radar sensing. The main idea is that in each iteration, we eliminate the ambiguity of the finer angle estimation based on the coarse angle estimation, and then refine the corresponding column of the sensing matrix.
\item {\bf We provide a framework of estimating the Doppler frequencies to support target speed measurement and payload data demodulation.} During the payload data transmission, a small amount of pilot signals are inserted between the adjacent data frames to estimate the Doppler frequencies (i.e., velocities) of targets and users. The Doppler estimation and compensation are vital for the data demodulation in communication-centric ISAC systems, which is, however, often sidestepped in the earlier works on ISAC.
\end{itemize}

\subsection{Notations}\label{S1.3}

Column vectors and matrices are denoted by lower- and upper-case boldface letters, respectively, while $(\cdot )^{*}$, $(\cdot )^{\rm T}$, $(\cdot )^{\rm H}$, and $(\cdot )^{\dag}$ denote the conjugate, transpose, conjugate transpose, and pseudo-inverse operators, respectively. $\mathbb{C}$ and $\mathbb{Z}$ are the sets of complex-valued numbers and integers, respectively. ${\bf 0}_{M\times N}$ and ${\bf I}_N$ are the $M\times N$ all-zero-element matrix and the $N\times N$ identity matrix, respectively. ${\cal CN}$ and ${\cal U}$ denote the complex Gaussian distribution and the uniform distribution, respectively. $[{\bf a}]_i$ denotes the $i$-th element of vector $\bf{a}$, while $[{\bf A}]_{i,j}$ represents the $i$-th row and $j$-th column element of matrix $\bf{A}$. $[{\bf A}]_{\cal I}$ ($[{\bf a}]_{\cal I}$) denotes the submatrix (subvector) consisting of the columns (elements) of $\bf{A}$ ($\bf{a}$) indexed by the ordered set $\cal I$. $\|\cdot \|_p$ and $\|\cdot \|_F$ are the $l_p$-norm and the Frobenius norm, respectively. $\otimes $ stands for the Kronecker product and ${\rm vec}(\cdot )$ is the vectorization operation according to the columns of the matrix. $\bmod (k,N)$, where $k,N \in \mathbb{Z}$, is the remainder after $k$ is divided by $N$. $\lceil x \rceil$ returns the smallest integer that is not smaller than $x$. ${\rm card} ({\cal I})$ is the cardinality of the set ${\cal I}$. $\Xi_N(x)$ is the $N$-order Dirichlet kernel function given by $\Xi_N(x) = \frac{\sin ( N x/2 )}{N \sin (x/2)}$ for $x \ne 2 k\pi$ and $\Xi_N( 2 k\pi ) = ( -1 )^{k (N - 1)}$ for $k \in \mathbb{Z}$. $\textsf{E}\{\cdot \}$ is the statistical expectation operator. The variables/notations associated with the radar system are overlined, e.g., $\overline{N}$, to distinguish them from the communication counterparts, e.g., $N$.

\section{System Description and Channel Model}\label{S2}

In this section, we present the generic model of a mmWave mMIMO ISAC system. The channels associated with the communications and radar are respectively modeled.

\subsection{System Model}\label{S2.1}

\begin{figure}[tp!]
\vspace*{-7mm}
\captionsetup{font={footnotesize,color={black}}, name = {Fig.}, singlelinecheck=off, labelsep = period}
\begin{center}
\subfigure[]{\includegraphics[width=3.2in]{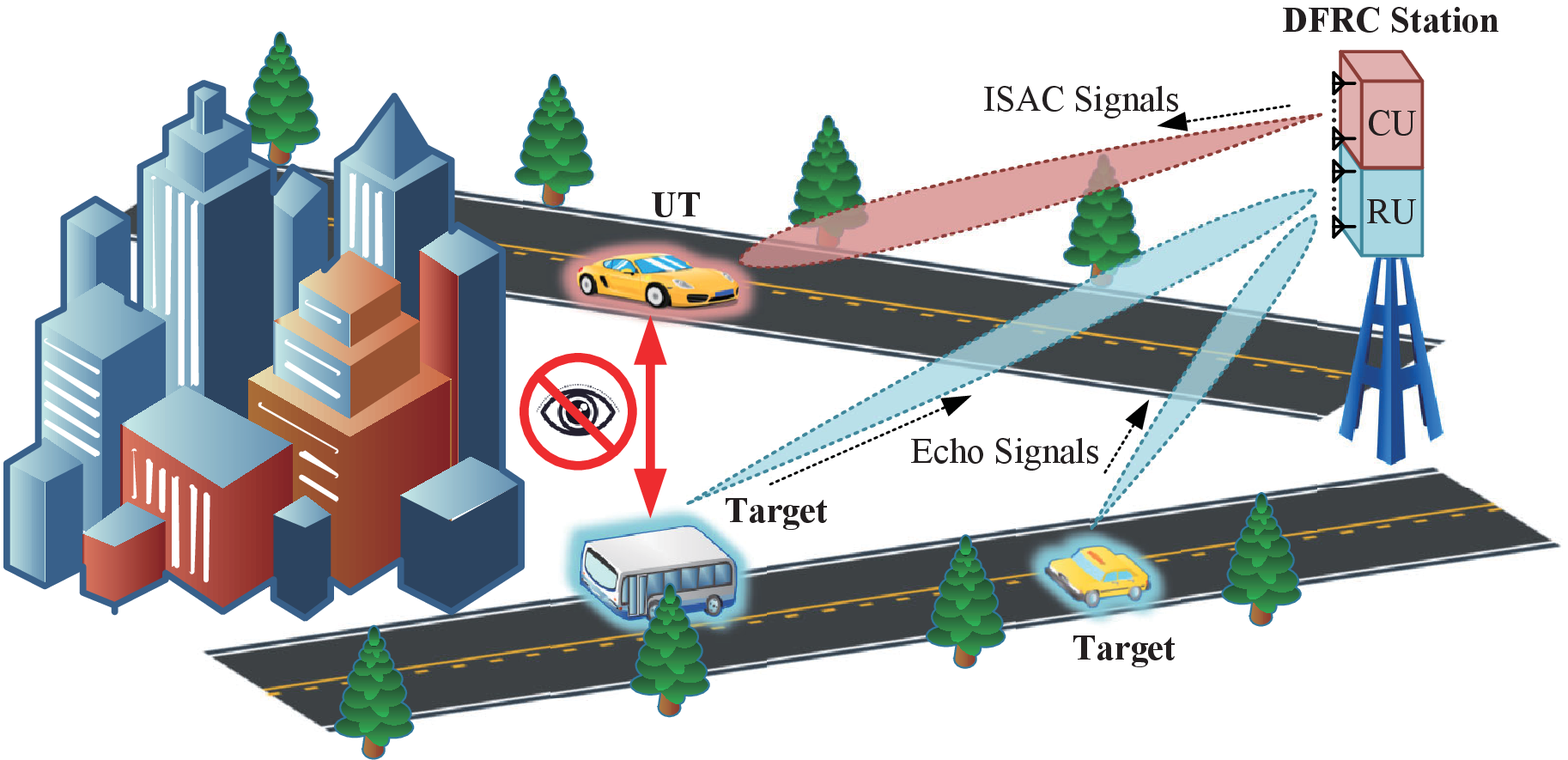}}
\\
\subfigure[]{\includegraphics[width=3in]{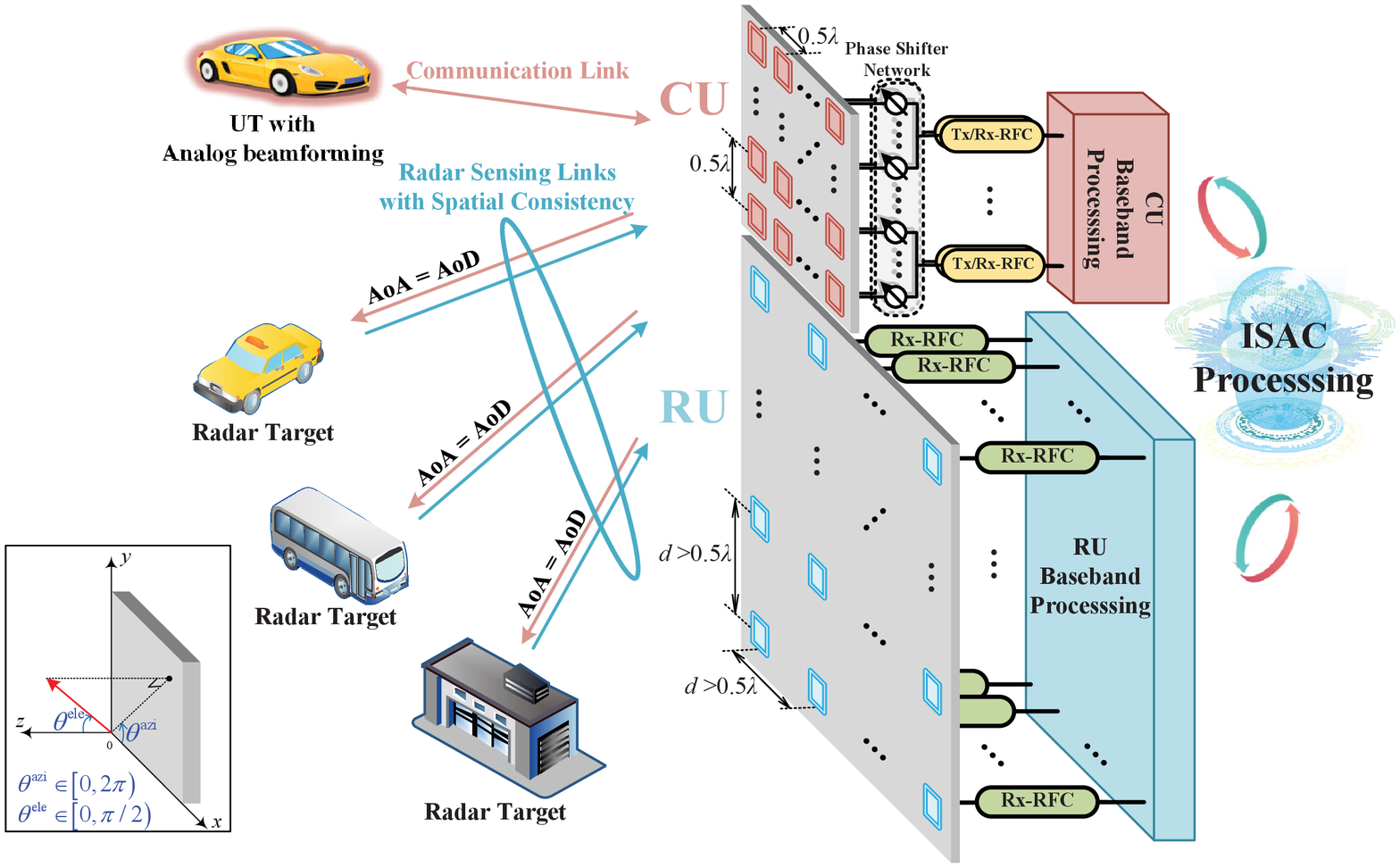}}
\hspace{6mm}
\subfigure[]{\includegraphics[width=1.5in]{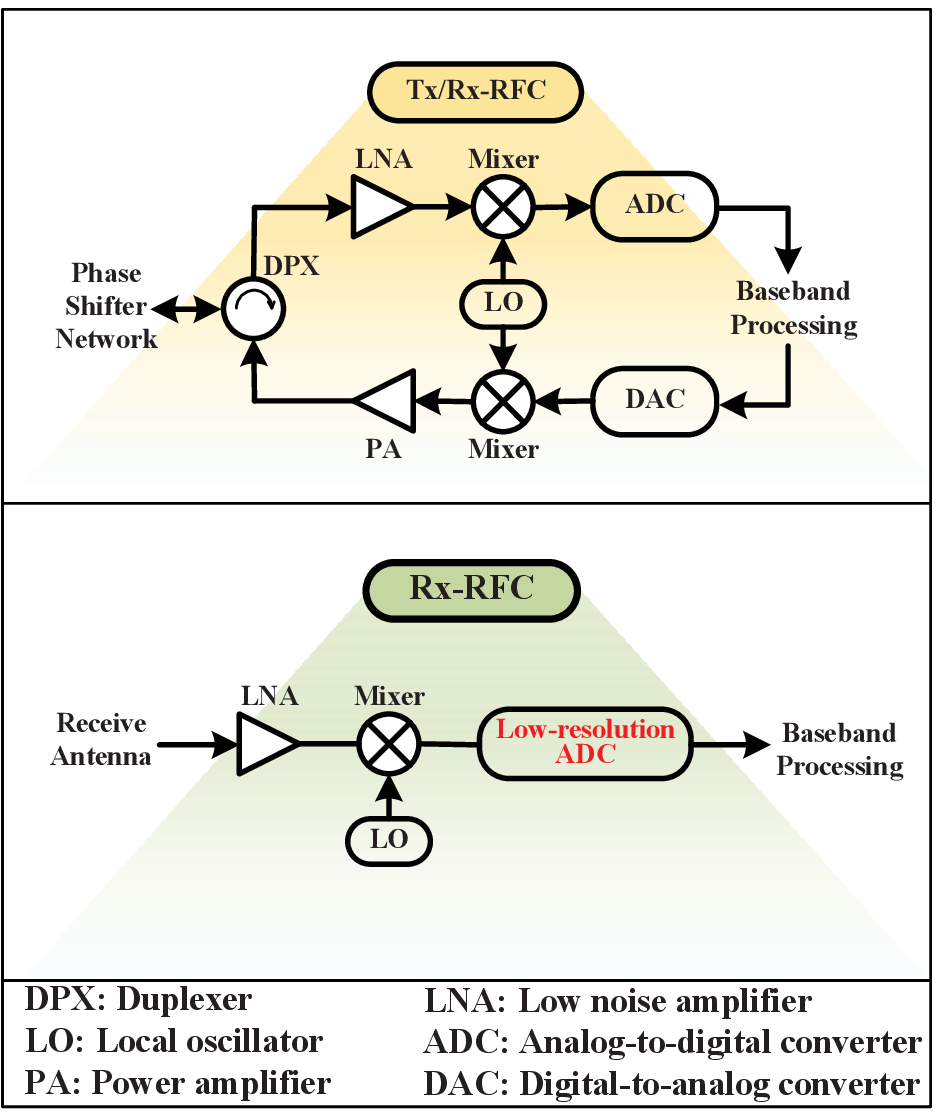}}
\end{center}
\vspace*{-7mm}
\caption{The model of an ISAC system assisted by mmWave mMIMO: (a)~An ISAC scenario where high-speed UTs are served, (b)~the hardware architectures of the CU and RU at the DFRC station, and (c)~Tx/Rx-RFC vs. Rx-RFC.}
\label{FigModel} 
\vspace{-5mm}
\end{figure}

Consider a mmWave mMIMO ISAC system shown in Fig. \ref{FigModel}\,(a), where a DFRC station serves multiple user-terminals (UTs) in a time division duplex (TDD) mode. This model can be applied to various ISAC scenarios. For instance, in vehicle-to-infrastructure (V2I) \cite{FLiuTWC} or vehicle-to-everything (V2X) systems \cite{MagaVT}, the road side units (RSUs) play the role of DFRC station, which needs to sense the environment and to provide vehicles the predictive alarm to avoid traffic accident. The center-carrier frequency of the system is $f_{\rm c}$ with the corresponding wavelength $\lambda$. Each UT is equipped with a uniform-planar-array (UPA) with $M = M_x \times M_y$ antennas, where $M_x$ and $M_y$ are the numbers of antennas along azimuth and elevation directions, respectively. Since UT is usually energy-constrained, analog beamforming technique is considered at each UT, i.e., there is only one RFC connected to $M$ antennas via $M$ phase shifters. The DFRC station consists of one CU and one RU, both equipped with UPAs. The CU is responsible for the communication tasks including information transmission and reception (Tx/Rx), while the RU only receives the echo signals for radar sensing. Compared to some previous works \cite{TcomSurvey,CM3} using a single array at the DFRC station for simultaneously transmitting and receiving, the proposed scheme provides a more hardware-feasible ISAC architecture with no need for full-duplex capability. The numbers of antennas at the CU and RU are $N = N_x \times N_y$ and $\overline N = \overline N_x \times \overline N_y$, respectively, where $N_x$ ($\overline N_x$) and $N_y$ ($\overline N_y$) are the numbers of antennas along azimuth and elevation directions, respectively. The arrays of the CU and RU are co-located and are parallel to each other so that they see the targets at the same propagation directions \cite{STAP,Tradeoff}.
We refer to this property as {\it spatial consistency}.

To achieve desired trade off of the power consumption, hardware cost, and system performance, we consider a dedicated DFRC architecture as shown in Fig.~\ref{FigModel}\,(b). At the CU, we employ the CSA with HBF architecture, where only $N_{\rm RF} \ll N$ Tx/Rx-RFCs are connected to $N$ antennas through a fully-connected phase shifters network, while at the RU, we consider the low-resolution ADC architecture \cite{MagaADC} with WSA. The antenna spacing of the CSA is half of the wavelength $d_{\rm cri} = 0.5{\lambda}$, while that of the WSA is set to $d > d_{\rm cri}$ so that $\overline{N}_x d > N_x d_{\rm cri}$ and $\overline{N}_y d > N_y d_{\rm cri}$, resulting in a larger spatial aperture. For each receive antenna at the RU, a dedicated Rx-RFC is deployed, as shown in Fig.~\ref{FigModel}\,(c). Although this architecture requires the number of Rx-RFCs to be equal to that of antennas at the RU, i.e., fully-digital receiver, it is feasible for the following reasons: (i)~compared to the Tx/Rx-RFC at the CU, no energy-hungry power amplifier (PA) is used in the Rx-RFC{\footnote{According to \cite{PALNA}, the power of a PA is 138 mW, while that of a low noise amplifier is 39 mW, both in the mmWave band.}}; (ii)~low resolution ADC is used in the Rx-RFC, which reduces power consumption and cost \cite{MagaADC}; and (iii)~a moderate number of antennas at the RU will be sufficient to achieve higher angular resolution and better radar sensing performance relying on WSA, and this point will be detailed in the sequel.

\subsection{Time-Varying Communications and Radar Channel Models}\label{S2.2}

In this subsection, we formulate the time-varying channel models for both the communications and radar. For the communications, we consider the Rician fading channel model with one line-of-sight (LoS) path and a few clustered non-LoS (NLoS) paths. Taking a downlink channel from the CU to a UT (user index is omitted for notation simplification) as an example, we have
\begin{align}\label{CommCh} 
{\bf H}(\tau ;t) =& {\bf H}_{\rm LoS}(t) p\left(\tau - \tau_{\rm LoS}(t) - \tau_{\rm p}\right) + \sum\limits_{c = 1}^{N_{\rm C}} \sum\limits_{l = 1}^{N_{\rm P}} {\bf H}_{c,l}(t) p\left(\tau - \tau_{c,l}(t) - \tau_{\rm p}\right)  ,
\end{align}
with
\begin{align} 
{\bf H}_{\rm LoS}(t) =& g_{\rm LoS}(t) {\bf a}_M\left( \theta_{\rm LoS}^{\rm azi}(t),\theta_{\rm LoS}^{\rm ele}( t) \right) {\bf a}_N^{\rm H}\left( \varphi_{\rm LoS}^{\rm azi}(t),\varphi_{\rm LoS}^{\rm ele}(t) \right) , \label{CommLoS} \\
{\bf H}_{c,l}(t) =& g_{c,l}(t) {\bf a}_M\left( \theta_{c,l}^{\rm azi}(t),\theta_{c,l}^{\rm ele}(t) \right) {\bf a}_N^{\rm H}\left( \varphi_{c,l}^{\rm azi}(t),\varphi_{c,l}^{\rm ele}(t) \right) . \label{CommNLoS} 
\end{align}
In \eqref{CommCh}, $t$ and $\tau$ are the time and delay variables, respectively, $N_{\rm C}$ and $N_{\rm P}$ are the number of clusters and the number of paths in each cluster, respectively, while $\tau_{\rm LoS}(t)$ and $\tau_{c,l}(t)$ are the delay-offsets of the LoS path and the $(c,l)$-th NLoS path, respectively. Furthermore, $p(\tau )$ is the pulse shaping filter function, and $\tau_{\rm p}$ is the single side duration of $p(\tau )$, i.e., $p(\tau )\! =\! 0$ when $| \tau | > \tau_{\rm p}$. In \eqref{CommLoS} and \eqref{CommNLoS}, $g_{\rm LoS}(t)$ and $g_{c,l}(t)$ are the channel coefficients of the LoS path and the $(c,l)$-th NLoS path, respectively, $\left\{ \theta_{\rm LoS}^{\rm azi}(t),\theta_{\rm LoS}^{\rm ele}(t) \right\}$ and $\left\{ \theta_{c,l}^{\rm azi}(t),\theta_{c,l}^{\rm ele}(t) \right\}$ are the angles-of-arrival (AoAs) of the LoS path and the $(c,l)$-th NLoS path, respectively, while $\left\{ \varphi_{\rm LoS}^{\rm azi}(t),\varphi_{\rm LoS}^{\rm ele}(t) \right\}$ and $\left\{ \varphi_{c,l}^{\rm azi}(t),\varphi_{c,l}^{\rm ele}(t) \right\}$ are the angles-of-departure (AoDs) of the LoS path and the $(c,l)$-th NLoS path, respectively. {Note that each AoA or AoD in \eqref{CommLoS} and \eqref{CommNLoS} contains the azimuth part (superscripted by ``azi'') ranging in $[ 0, \, 2\pi )$ and the elevation part (superscripted by ``ele'') ranging in $[ 0, \, \pi /2 )$ (see Fig.~\ref{FigModel}\,(b)).} The steering vector ${\bf a}_N\left( \varphi^{\rm azi},\varphi^{\rm ele} \right)\! \in\! \mathbb{C}^{N \times 1}$ is given by
\begin{align}\label{CSAStrVec}  
{\bf a}_N\left( \varphi^{\rm azi},\varphi^{\rm ele} \right) =& {\bf a}\left( \mu_{\varphi} ; N_x \right) \otimes {\bf a}\left( \nu_{\varphi} ; N_y \right) ,
\end{align}
where $\mu_{\varphi} = \cos \varphi^{\rm azi}\sin \varphi^{\rm ele}$, $\nu_{\varphi} = \sin \varphi^{\rm azi} \sin \varphi^{\rm ele}$, and
\begin{align} 
{\bf a}\left( \mu_{\varphi} ; N_x \right) =& \frac{1}{\sqrt{N_x}} \left[ 1 ~ e^{-\textsf{j}\pi\mu_{\varphi}} \cdots e^{-\textsf{j}\pi ( N_x - 1 ) \mu_{\varphi}} \right]^{\rm T} \in \mathbb{C}^{N_x \times 1} , \label{CSAStrVecPart} \\
{\bf a}\left( \nu_{\varphi} ; N_y \right) =& \frac{1}{\sqrt{N_y}} \left[ 1 ~ e^{-\textsf{j}\pi\nu_{\varphi}} \cdots e^{-\textsf{j}\pi ( N_y - 1 ) \nu_{\varphi}} \right]^{\rm T} \in \mathbb{C}^{N_y \times 1} . \label{CSAStrHorPart}
\end{align}
Furthermore, ${\bf a}_M\left( \theta^{\rm azi},\theta^{\rm ele} \right)\! =\! {\bf a}\left( \mu_{\theta} ; M_x \right) \otimes {\bf a}\left( \nu_{\theta} ; M_y \right)$ can be formulated similarly to \eqref{CSAStrVecPart} and \eqref{CSAStrHorPart}.

On the other hand, the channel for radar sensing (from the CU to the target and then back to the RU) can be formulated as
\begin{align}\label{RadarCh} 
\overline{\bf{H}}( \tau ;t ) =& \sum\limits_{c = 1}^{\overline{N}_{\rm{C}}} \sum\limits_{l = 1}^{\overline{N}_{\rm{P}}} \overline{\bf{H}}_{c,l}(t) p\left( \tau - \overline{\tau}_{c,l}(t) - \tau_{\rm{p}} \right) ,
\end{align}
where
\vspace{-1mm}
\begin{align}\label{RadarNLoS} 
\overline{\bf{H}}_{c,l}(t) =& \overline{g}_{c,l}(t) \overline{\bf{a}}_{\overline{N}}\left( \overline{\theta}_{c,l}^{\rm{azi}}(t),\overline{\theta}_{c,l}^{\rm{ele}}(t) \right) {\bf{a}}_{{N}}^{\rm H}\left(\overline{\theta}_{c,l}^{\rm{azi}}(t),\overline{\theta}_{c,l}^{\rm{ele}}(t) \right) .
\end{align}
In \eqref{RadarCh}, $\overline{N}_{\rm{C}}$ is the number of radar targets of interest, $\overline{N}_{\rm{P}}$ is the number of resolvable paths induced by a target, and $\overline{\tau}_{c,l}(t)$ is the delay of the $l$-th echo signal of the $c$-th target. In \eqref{RadarNLoS}, $\overline{g}_{c,l}(t)$ is the coefficient of the $l$-th path of the $c$-th target, which accounts for the free space propagation loss and the radar cross section (RCS) of the target, while $\left\{ \overline {\theta}_{c,l}^{\rm{azi}}(t),\overline{\theta}_{c,l}^{\rm{ele}}(t) \right\}$ represents the angle {(also including the azimuth part and elevation part like those in \eqref{CommLoS} and \eqref{CommNLoS})} of the $l$-th path of the $c$-th target. Note that due to the aforementioned spatial consistency, $\left\{ \overline{\theta}_{c,l}^{\rm{azi}}(t),\overline{\theta}_{c,l}^{\rm{ele}}(t) \right\}$ is shared by the transmitter and the receiver in \eqref{RadarNLoS}. Also note that we ignore the multi-hop signals in \eqref{RadarNLoS}. The steering vector of WSA $\overline{\bf{a}}_{\overline{N}}\left( \overline{\theta }^{\rm{azi}},\overline{\theta}^{\rm{ele}} \right) \in \mathbb{C}^{\overline{N} \times 1}$ can be formulated with the parametrized antenna spacing $d$ as
\begin{align}\label{WSAStrVec} 
\overline{\bf{a}}_{\overline{N}}\left( \overline{\theta}^{\rm{azi}},\overline{\theta}^{\rm{ele}} \right) =& \overline{\bf{a}}\left( \overline{\mu}_{\overline{\theta}}; \overline{N}_x \right) \otimes \overline{\bf{a}}\left( \overline{\nu}_{\overline{\theta}}; \overline{N}_y \right) ,
\end{align}
where $\overline{\mu}_{\overline{\theta}} = \cos \overline{\theta}^{\rm azi} \sin \overline{\theta}^{\rm ele}$, $\overline{\nu}_{\overline{\theta}} = \sin \overline{\theta}^{\rm azi} \sin \overline{\theta}^{\rm ele}$, and
\begin{align} 
\overline{\bf{a}}\left(\overline{\mu}_{\overline{\theta}}; \overline{N}_x \right) =& \frac{1}{\sqrt{\overline{N}_x}} \left[ 1 ~ e^{-\textsf{j}\frac{2 \pi d}{\lambda} \overline{\mu}_{\overline{\theta}}} \cdots e^{-\textsf{j}\frac{2 \pi d}{\lambda} \left( \overline{N}_x - 1 \right) \overline{\mu}_{\overline{\theta}}} \right]^{\rm T} , \label{WSAStrVecPart} \\
\overline{\bf{a}}\left(\overline{\nu}_{\overline{\theta}}; \overline{N}_y \right) =& \frac{1}{\sqrt{\overline{N}_y}} \left[ 1 ~ e^{-\textsf{j}\frac{2 \pi d}{\lambda} \overline{\nu}_{\overline{\theta}}} \cdots e^{-\textsf{j}\frac{2 \pi d}{\lambda} \left( \overline{N}_y - 1 \right) \overline{\nu}_{\overline{\theta}}} \right]^{\rm T} \label{WSAStrHriPart} .
\end{align}

It is worth noting that by taking into account the high angular resolution offered by mmWave mMIMO, we formulate each target as a cluster with multiple resolvable paths in \eqref{RadarCh}. Compared with some previous literature \cite{TcomSurvey,AZhang,CSJRC1,CSJRC2} assuming that each target contributes only a single path (i.e., point target, corresponding to the case of $\overline{N}_{\rm p} = 1$), our channel model \eqref{RadarCh} facilitates target imaging through the mmWave RF signals \cite{AccessImaging}. In other words, by estimating the channels in the angular domain and identifying the shape (geometric information) of each cluster, the imaging information of the targets can be obtained, even when the targets are covered by materials that can easily block the visible light and laser signals \cite{AccessImaging}. This may extend the scope of target imaging when traditional imaging methods, e.g., camera or LiDAR \cite{MagaVT}, fail, and thus may improve the efficiency, reliability and safety in future ISAC scenarios.

\section{Frame Structure and Waveform Design}\label{S3}

In this section, we design a transmission frame structure for the ISAC system, which considers different timescales. Moreover, we formulate the input-output signal models in the ISAC system, and propose a waveform design, which not only guarantees the performance of CS-based algorithms but also meets the practical hardware constraints imposed by the HBF architecture.

\subsection{Frame Structure Design}\label{S3.1}

Although the channel models \eqref{CommCh} and \eqref{RadarCh} are time-varying, some or all of their parameters may be reasonably assumed to be time-invariant under different timescales. More specifically,
\begin{itemize}
\item In a small timescale that is much shorter than the channel coherence time\footnote{The channel coherence time can be predicted based on Doppler spread. According to \cite{Book,GaoTSP}, the channel coherence time is $T_{\rm coh} \approx \sqrt{9/(16\pi f_{\max}^2)}$, where $f_{\rm max}$ is the maximum Doppler frequency shift.
}, the whole channel can be assumed to be time-invariant. Therefore, we can drop the time index $t$ in \eqref{CommCh} and \eqref{RadarCh}, i.e., ${\bf{H}}( \tau ;t) = {\bf{H}}( \tau )$ and $\overline{\bf{H}}( \tau ;t ) = \overline{\bf{H}}( \tau )$, in this case.
\item In a moderate timescale, the velocities and the positions of scatters or targets are relatively stable.  Hence, the delays, angles and Doppler frequencies of scatters or targets can be assumed to be time-invariant, as their changes may be regarded as negligible. Yet, the channel coefficients $\left\{ g_{\rm{LoS}}(t),g_{c,l}(t),\overline{g }_{c,l}( t) \right\}$ may vary significantly due to the Doppler effect.
\item In a large timescale, the environment may experience dramatic changes, and the channels become significantly different from the previously acquired channel information. A new CE and radar sensing stage should be carried out to capture the changes.
\end{itemize}

\begin{figure}[tp!]
\vspace*{-4mm}
\captionsetup{font={footnotesize,color={black}}, name = {Fig.}, singlelinecheck=off, labelsep = period}
\begin{center}
\includegraphics[width=5.5in]{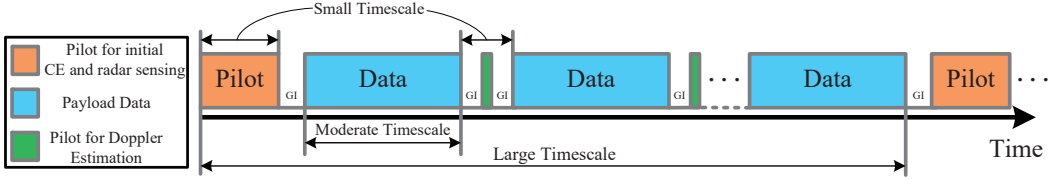}
\end{center}
\vspace*{-8mm}
\caption{The proposed transmission frame structure of the ISAC system. The guard interval (GI) is inserted between the pilot signals and payload data to avoid inter-frame interference.}
\label{FigFrmStru} 
\vspace{-7mm}
\end{figure}

Based on the analysis above, we illustrate our proposed transmission frame structure in Fig.~\ref{FigFrmStru}.
In the initial joint CE and radar sensing stage, the CU transmits the pilot signals for both CE and radar sensing, i.e., ISAC signals. By leveraging the received pilot and the echo signals, the UT and the RU conduct the CE and radar sensing, respectively. Then, the CU and the UT formulate the transmit and receive beamformer based on the results from CE and radar sensing in order to guarantee the quality of the following payload data communications and target tracking. Besides, pilot signals with a very short duration are inserted between two adjacent payload data blocks to estimate the potential Doppler components of the served UTs or the targets of interest.

\subsection{Problem Formulation and Waveform Design}\label{S3.2}

We focus on the initial joint CE and radar sensing stage in Fig.~\ref{FigFrmStru}. Note that the duration of this stage or pilot duration is much shorter than the channel coherence time so that the channels can be seen as time-invariant, i.e., the time index $t$ can be dropped. Let $L\! =\! \left\lceil \frac{\max\limits_{c,l} \left\{ \tau_{c,l},\overline{\tau}_{c,l} \right\} + 2 \tau_{\rm{p}} }{T_{\rm{s}}} \right\rceil\! +\! 1$ be the maximum delay spread (in samples) where $T_{\rm s}$ is the sampling period of the system. The channel impulse response (CIR) of the communication channel can be written as
\begin{align}\label{CIR} 
{\bf{H}}_l =& \left\{ \begin{array}{cl}
 {\bf{H}}\left( l T_{\rm{s}} \right) ,  & l = 0,1,\cdots ,L-1 , \\
 {\bf{0}}_{M \times N} , & {\rm{Others}} .
\end{array} \right.
\end{align}
The CIR of the radar channel $\overline{\bf{H}}_l$, $l=0,1,\cdots ,L-1$, can be  formulated similarly.

Consider that the CU transmits the pilot signals with length $P$, i.e., ${\bf{p}}_p \in \mathbb{C}^{N \times 1}$, $p=0,1,\cdots,P-1$, which are known to both the transmitters and receivers. We define ${\bf{p}}_p = {\bf{0}}_{N \times 1}$ for $p < 0$ and $p \ge P$, since a sufficiently long zero guard interval (GI) should be inserted between the pilot signals and payload data to avoid the inter-frame interference and to provide enough time for reconfiguring the RF circuits \cite{HeathJSAC}. Thus the pilot signals received by the RU and the UT with noise in the $n$-th time slot can be obtained respectively based on linear convolution as 
\begin{align} 
\overline{\bf{y}}_n =& \textsf{Q} \left\{ \sum\limits_{l = 0}^{L-1} \overline{\bf{H}}_l {\bf{p}}_{n - l} + \overline{\bf{n}}_n \right\}  , \label{RURxS} \\
y_n =& {\bf w}^{\rm H}_n \sum\limits_{l = 0}^{L-1} {\bf{H}}_l {\bf{p}}_{n - l} + {\bf w}^{\rm H}_n {\bf n}_n , \label{UTRxS}
\end{align}
respectively, where $\textsf{Q}\{ \cdot \}$ is the quantization function caused by the low-resolution ADCs at the RU{\footnote{We adopt high-resolution ADCs at the UT for reliable communications, so the quantization function is ignored in \eqref{UTRxS}.}}, $\overline{\bf{n}}_n \sim {\cal{CN}}\big({\bf 0}_{\overline{N} \times 1},\sigma_{\rm n}^2{\bf{I}}_{\overline{N}}\big)$ and ${\bf n}_n \sim {\cal{CN}}\big({\bf 0}_{M \times 1},\sigma_{\rm n}^2{\bf{I}}_M\big)$ are the additive white Gaussian noise (AWGN) vectors at the RU and UT, respectively, while ${{\bf{w}}_n} \in \mathbb{C}^{M \times 1}$ is the analog weight vector at the UT. Note that according to the property of linear convolution, the length of the sequence $\overline{\bf{y}}_n$ (or $y_n$) is $Q {\buildrel \Delta \over =} P+L-1$, i.e., $\overline{\bf{y}}_n = \bf{0}$ (or $y_n = 0$) if $n < 0$ or $n \ge Q$.

At the RU, by collecting all the measurements $\left\{ \overline{\bf{y}}_n \right\}_{n = 0}^{Q-1}$, we have
\begin{align}\label{RadCSpro} 
\overline{\bf{Y}} =& \left[ \overline{\bf{y}}_0 ~ \overline{\bf{y}}_1 \cdots \overline{\bf{y}}_{Q-1} \right] = \textsf{Q} \left\{ \overline{\bf{H}}_{\rm{SD}} \overline{\bf{\Phi}} + \overline{\bf{N}} \right\} ,
\end{align}
where $\overline{\bf{H}}_{\rm{SD}} = \left[ \overline{\bf{H}}_0 ~ \overline{\bf{H}}_1 \cdots \overline{\bf{H}}_{L - 1} \right] \in \mathbb{C}^{\overline{N} \times LN}$ is the effective CIR in the spatial-delay (SD) domains, $\overline{\bf{N}} = \left[ \overline{\bf{n}}_0 ~ \overline{\bf{n}}_1 \cdots \overline{\bf{n}}_{Q-1} \right] \in \mathbb{C}^{\overline{N} \times Q}$ is the effective noise matrix, and $\overline{\bf{\Phi}} \in \mathbb{C}^{LN \times Q}$ is the known effective measurement matrix with the block-Toeplitz property. Specifically, the $q$-th column of $\overline{\bf{\Phi}}$, $1 \le q \le Q$, can be expressed as
\begin{align}\label{PhiRad} 
\left[\overline{\bm{\Phi}}\right]_{\{q\}} =& {\left[ {{\bf{p}}_{q - 1}^{\rm{T}} ~ {\bf{p}}_{q - 2}^{\rm{T}} \cdots {\bf{p}}_{q - L}^{\rm{T}}} \right]^{\rm{T}}} ,
\end{align}
with ${\bf{p}}_p = {\bf{0}}_{N \times 1}$ for $p < 0$ and $p \ge P$. Similarly, at the UT, we collect $\left\{ y_n \right\}_{n = 0}^{Q-1}$ to obtain
\begin{align}\label{ComCSpro} 
{\bf{y}} =& \left[ y_0 ~ y_1 \cdots y_{Q-1} \right]^{\rm T} = {\bf{\Phi}}\, {\rm{vec}}\left( {\bf{H}}_{\rm{SD}} \right) + {\bf{n}} ,
\end{align}
where ${\bf{H}}_{\rm{SD}}\! =\! \left[ {\bf{H}}_0 ~ {\bf{H}}_1\cdots {\bf{H}}_{L - 1} \right]\! \in\! \mathbb{C}^{M \times LN}$ and ${\bf{n}}\! =\! \left[ {\bf w}_0^{\rm H} {\bf n}_0 ~ {\bf w}_1^{\rm H} {\bf n}_1\cdots {\bf w}_{Q-1}^{\rm H} {\bf n}_{Q-1} \right]^{\rm T}\! \in\! \mathbb{C}^{Q \times 1}$, while
\begin{align}\label{PhiCom} 
{\bm \Phi}  = \left[ {{\bf{b}}_1^{} \otimes {\bf{w}}_0^* ~~ {\bf{b}}_2^{} \otimes {\bf{w}}_1^* ~ \cdots ~ {\bf{b}}_Q^{} \otimes {\bf{w}}_{Q - 1}^*} \right]_{}^{\rm{T}} \in \mathbb{C}^{Q \times LMN} ,
\end{align}
in which ${\bf b}_q\! =\! \left[ \overline{\bf{\Phi}} \right]_{\{q\}}$ as formulated in \eqref{PhiRad}. Note that the radar CIR in the spatial-delay domain $\overline{\bf{H}}_{\rm{SD}}$ contains useful information (e.g., initial positions and shapes) of the targets. In this paper, we focus on estimating $\overline{\bf{H}}_{\rm{SD}}$ accurately for the initial radar sensing. The further data processing for extracting the features of the targets from $\overline{\bf{H}}_{\rm{SD}}$ however is beyond the scope of this paper. Interested readers can see the examples in \cite{AccessImaging} for more information.

From \eqref{RadCSpro} and \eqref{ComCSpro}, it is seen that ${\bf{\overline H}}_{{\rm{SD}}}^{}$ and ${\bf{H}}_{{\rm{SD}}}^{}$ are observed via the linear systems (temporarily ignoring $\textsf{Q}\{\cdot\}$ for simplicity) with additive noise, which is a canonical form of CS problems \cite{GaoCM}. By leveraging channel sparsity and sophisticated CS-based algorithms, high-accuracy channel recovery can be achieved even with under-determined measurements (i.e., $\overline{N}Q < LN\overline{N}$ and $Q < LMN$), and we will discuss this point in the next subsection. Given that the accuracy of CS-based algorithms depends heavily on the structure of measurement matrices (i.e., ${\bf \Phi}$ and ${\bf \overline \Phi}$), we propose a pilot waveform design scheme by taking into account both the CS theories and hardware constraints. The pilot waveform ${{\bf{p}}_p}$ can be expressed as a product of the analog precoder ${{\bf{F}}_p} \in \mathbb{C}^{N \times N_{\rm RF}}$ and the baseband pilot symbols ${{\bf{s}}_p} \in \mathbb{C}^{N_{\rm RF} \times 1}$ as
\begin{align}\label{PilotSyb} 
{{\bf{p}}_p} =& {\bf{F}}_p {\bf{s}}_p ,
\end{align}
where the elements of ${\bf{F}}_p$ satisfy $\left| \left[ {\bf{F}}_p \right]_{i,j} \right|^2 = N^{-1}$, $1 \le i \le N$, $1 \le j \le N_{\rm RF}$, owing to the constant-modulus property of phase shifters. Note that the elements of analog combiner ${\bf w}_n$ also satisfy the constant-modulus constraint.

\begin{remark}\label{REM1}
Since the under-determined estimation problems are considered in \eqref{RadCSpro} and \eqref{ComCSpro}, the orthogonal waveform design which minimizes the CRB of parameter estimation \cite{TcomSurvey} is not applicable. 
Indeed, the pilot waveform should be carefully designed under the CS theoretic framework and hardware constraints.
Particularly, in HBF architecture, the switch of phase shifter will take non-negligible reconfiguring time \cite{HeathJSAC}. During this time, the phase value of each phase shifter is uncertain and thus the transmit/receive pilot signals generated by analog phase shifters in this period are unpredictable. Thus it is impractical to use two different analog precoders (combiners) for two adjacent transmit (receive) pilot signals, as done in \cite{TcomSurvey}. Instead, sufficient idle time between switching two different analog precoders or combiners should be reserved. At the same time, the invalid receive pilot signals caused by the uncertain analog combiners should be removed from the measurements at the UT.
\end{remark}

\begin{figure}[tp]
\captionsetup{font={footnotesize,color={black}}, name = {Fig.}, singlelinecheck=off, labelsep = period}
\begin{center}
\includegraphics[width=4.5in]{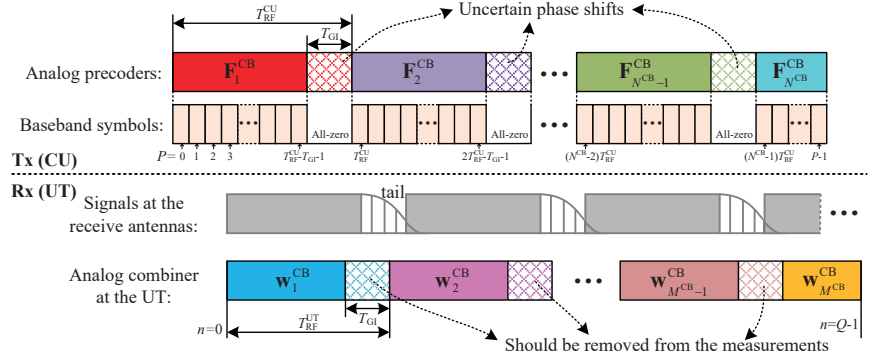}
\end{center}
\vspace*{-5mm}
\caption{The proposed waveform design which considers both the pilot diversity and the hardware feasibility. Since the tail parts at the receive antennas are also valid measurements of channels, we do not require perfect alignment between the GIs of the transmitter and receiver.}
\label{FigWFdesign} 
\vspace{-7mm}
\end{figure}

Taking both the pilot diversity and the hardware feasibility into account, the proposed pilot waveform design can be summarized as follows. First, we define $T_{\rm RF}^{\rm CU}$ and $T_{\rm RF}^{\rm UT}$ as the durations of applying the same precoder and combiner at the CU and UT, respectively, and $T_{\rm GI} < \min \{T_{\rm RF}^{\rm CU},T_{\rm RF}^{\rm UT} \}$ as the required GI for reconfiguring RF circuits
(all in samples). Next, we induce a codebook of the precoder $\left\{ {\bf{F}}_n^{\rm{CB}} \in \mathbb{C}^{N \times N_{\rm RF}} \right\}_{n = 1}^{N^{\rm{CB}}}$ and a codebook of the combiner $\left\{ {\bf{w}}_n^{\rm{CB}} \in \mathbb{C}^{M \times 1} \right\}_{n = 1}^{M^{\rm{CB}}}$, where $N^{\rm{CB}} = \left\lceil P/T^{\rm CU}_{\rm{RF}} \right\rceil$ and $M^{\rm{CB}} = \left\lceil Q/T^{\rm UT}_{\rm{RF}} \right\rceil$. Then, the analog precoder ${\bf F}_p$, $0 \le p < P$, and the analog combiner ${\bf w}_n$, $0 \le n < Q$, are designed respectively as
\begin{align} 
{\bf{F}}_p =& \left\{ \begin{array}{cl}
 {\bf{F}}_{\left\lceil (p + 1)/T^{\rm CU}_{\rm{RF}} \right\rceil}^{\rm{CB}} , & \bmod \left( p, T^{\rm CU}_{\rm{RF}} \right) <  \left( T^{\rm CU}_{\rm{RF}} - T_{\rm{GI}} \right) \text{ or } {\left\lceil (p + 1)/T^{\rm CU}_{\rm{RF}} \right\rceil} = N^{\rm CB}, \\
 {\rm{uncertain}} , & {\rm{otherwise}} ,
\end{array} \right.  \label{design1} \\
{\bf{w}}_n =& \left\{ \begin{array}{cl}
 {\bf{w}}_{\left\lceil (n + 1)/T^{\rm UT}_{\rm{RF}} \right\rceil}^{\rm{CB}} , & \bmod \left( n, T^{\rm UT}_{\rm{RF}} \right) < \left( T^{\rm UT}_{\rm{RF}} - T_{\rm{GI}} \right) \text{ or } {\left\lceil (n + 1)/T^{\rm UT}_{\rm{RF}} \right\rceil} = M^{\rm CB}, \\
 {\rm{uncertain}} , & {\rm{otherwise}} .
\end{array} \right. \label{design2}
\end{align}
This design scheme is intuitively explained in Fig.~\ref{FigWFdesign}. At the CU, the pilot signals are divided into $N^{\rm CB}$ sub-frames, and the signals in each sub-frame share the same analog precoder. In the last $T_{\rm GI}$ samples of the first $(N^{\rm CB}\! -\! 1)$ sub-frames\footnote{The last sub-frame does not need the extra reconfiguring time, since the RF circuits can be reconfigured during the GI inserted before the next frame (see Fig.~\ref{FigFrmStru}).}, the precoder is switched to a different one, which results in uncertain values of phase shifters. During this reconfiguring time, zero baseband signals are transmitted, i.e., the actual transmit pilot signals are also zero:
\begin{align}\label{ZeroGI} 
{\bf s}_p =& {\bf 0}_{N_{\rm RF} \times 1} \text{ when } \bmod \left( p, T^{\rm CU}_{\rm{RF}} \right) \ge \left( T^{\rm CU}_{\rm{RF}} - T_{\rm{GI}} \right) \text{ and } {\left\lceil (p + 1)/T^{\rm CU}_{\rm{RF}} \right\rceil} \neq N^{\rm CB}.
\end{align}
This all-zero GI does not mean that there is no pilot signal received at the receivers, since $T_{\rm GI} \ll L$ and thus the tail part of the previous pilot signals can be received by the receivers contributing to the effective measurements for CE{\footnote{Due to the existence of tail part, we do not require perfect alignment between the GIs of the transmitter and receiver, However, the perfect alignment will render the best performance of CE, and it can be guaranteed by the reliable frame synchronization based on preambles with good auto-correlation property (e.g., Zadoff-Chu sequence).}}. Similarly, at the UT, $M^{\rm CB}$ analog combiners are assigned to $M^{\rm CB}$ sub-frames of receive pilot signals, and a $T_{\rm GI}$-length GI is inserted into each sub-frame (except for the last sub-frame) in order to reconfigure the phase shifters. Moreover, given the uncertain analog combiners, the receive pilot signals during each GI should be removed from the measurements in \eqref{ComCSpro}, which yields
\begin{align}\label{CommCSFinal} 
{\bf{y}}_{\rm{valid}} =& {\bf{\Phi}}_{\rm{valid}} \, {\rm{vec}}\left( {\bf{H}}_{\rm{SD}} \right) + {\bf{n}}_{\rm{valid}} ,
\end{align}
where ${\bf{y}}_{\rm valid} = \left[ {\bf{y}} \right]_{{\cal I}_{\rm{valid}}}$, ${\bf{n}}_{\rm{valid}} = \left[ {\bf{n}} \right]_{{\cal I}_{\rm{valid}}}$, ${\bf{\Phi}}_{\rm{valid}} = \left[ {\bf{\Phi}} \right]_{{\cal I}_{\rm{valid}}}$, and the ordered set ${\cal I}_{\rm{valid}} = \left\{ n + 1 \left| \, 0 \le n < Q , \, \bmod \left( n, T^{\rm UT}_{\rm{RF}} \right) < \left( T^{\rm UT}_{\rm{RF}} - T_{\rm{GI}} \right) \text{ or } {\left\lceil (n + 1)/T^{\rm UT}_{\rm{RF}} \right\rceil} = M^{\rm CB} \right. \right\}$. In the next subsection, we will focus on recovering the channels in \eqref{RadCSpro} and \eqref{CommCSFinal} via CS-based algorithms.
Note that by setting different $N^{\rm CB}$ ($M^{\rm CB}$), higher pilot diversity can be achieved under the practical HBF architecture, and it is expected to benefit the CE and radar sensing from the perspective of CS \cite{GaoCL}. Furthermore, compared with the previous work \cite{TcomSurvey} which designs different precoders for each symbol, the proposed pilot waveform design imposes dramatically lower storage requirements for the pre-defined codebooks, since we have $N^{\rm CB} \ll P$ and $M^{\rm CB} \ll Q$.

The specific design of the codebooks is detailed as follows. The baseband pilot symbols ${\bf s}_p$ are randomly drawn from the symbol set with normalized transmit power. In this paper, we consider binary phase shift keying (BPSK) symbols for the initial CE and radar sensing stage, and thus each element in those pilot symbols ${\bf s}_p \neq {\bf 0}_{N_{\rm RF} \times 1}$ is equiprobably drawn from the set $\left\{ -\sqrt{ \frac{P_{\rm DL}}{N_{\rm{RF}}} },\, \sqrt{ \frac{P_{\rm DL}}{N_{\rm{RF}}} } \right\}$, where $P_{\rm DL}$ is the total transmit power of downlink pilot signals. In addition, randomized phase shifts \cite{GaoCL} are employed for the precoder and combiner codebooks, i.e., each element of ${\bf F}^{\rm CB}_p$ (${\bf w}^{\rm CB}_n$) is set to $\frac{1}{\sqrt{N}}e^{\textsf{j}\phi}$ ($\frac{1}{\sqrt{M}}e^{\textsf{j}\phi}$) with the random variable $\phi \sim {\cal{U}}\left[ 0,\, 2\pi  \right)$.

\section{Compressive Sensing For ISAC and Doppler Estimation}\label{S4}

In this section, we formulate the CE and radar sensing as the sparse signal recovery problems with the designed dictionaries for WSA. A CS-based algorithm is proposed by taking into account the spatial consistency in order to solve the problems and obtain higher angular resolution. We also provide a framework for estimating the Doppler frequencies of UTs/targets. 

\begin{figure}[bp!]
\vspace*{-7mm}
\captionsetup{font={footnotesize}, name = {Fig.}, singlelinecheck=off, labelsep = period}
\begin{center}
\includegraphics[width=5.5in]{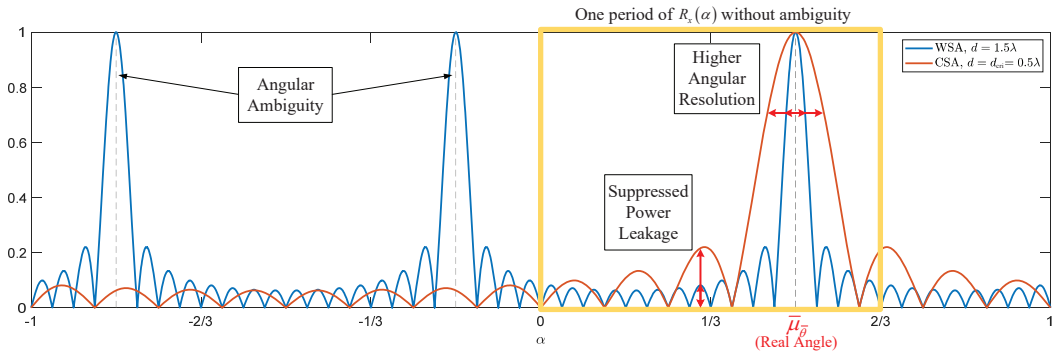}
\end{center}
\vspace*{-5mm}
\caption{An example of the correlation function $R_x\left( {\alpha} \right)$ associated with the WSA, where $\overline{N}_x = 16$, $d = 1.5\lambda$ and $\overline{\mu}_{\overline \theta} = 0.5$, in comparison with the case of the CSA ($d = d_{\rm cri} = 0.5\lambda$).}
\label{FigCorFun} 
\vspace{-5mm}
\end{figure}

\subsection{Dictionary Design}\label{S4.1}

As mentioned previously, the same channel angles are experienced by the CU and RU due to spatial consistency. Also, different angular resolutions are imposed by the CSA at the CU and the WSA at the RU. These two properties motivate us to design a dedicated CS-based algorithm to obtain robust radar sensing performance against the angular ambiguity. We first investigate the angular-domain channel associated with WSA. Without loss of generality, we focus on a single steering vector of WSA ${\overline{\bf a}}_{\overline{N}}\left( \overline{\theta}^{\rm azi},\overline{\theta}^{\rm ele} \right)$ given in \eqref{WSAStrVec}. The analysis can be readily extended to multi-path scenarios as in \eqref{RadarCh}. A direct method for estimating angle $\{ \overline{\theta}^{\rm azi},\overline{\theta}^{\rm ele} \}$ is to find the peak of the correlation function between ${\overline{\bf a}}_{\overline{N}}\left( \overline{\theta}^{\rm azi},\overline{\theta}^{\rm ele} \right)$ and a probing vector ${\overline{\bf a}}_{\overline{N}}\left( x,y \right)$. By defining $\alpha = \cos x \sin y$ and $\beta = \sin x \sin y$, the correlation function can be expressed as \cite{MyTcom}
\begin{align}\label{CorFun} 
R( \alpha ,\beta ) =& \left| \overline{\bf{a}}_{\bar N}^{\rm H}( x,y ) \overline{\bf{a}}_{\bar N}\left( \overline{\theta}^{\rm{azi}},\overline{\theta}^{\rm{ele}} \right) \right| 
 = \left| \Xi_{\bar{N}_x}\left( \frac{2\pi d}{\lambda}\left( \alpha - \overline{\mu}_{\overline{\theta}} \right) \right) \right| \times \left| \Xi_{\bar{N}_y}\left( \frac{2\pi d}{\lambda}\left( \beta - \overline{\nu}_{\overline {\theta}} \right) \right) \right| .
\end{align}
We refer to $\overline{\mu}_{\overline{\theta}}$ and $\overline{\nu}_{\overline{\theta}}$ as the virtual azimuth angle and the virtual elevation angle, respectively, while $R( \alpha ,\beta )$ can be viewed as the absolute value of the angular-domain channel. Note that there is a one-to-one mapping between $\{ \overline{\mu}_{\overline{\theta}},\overline{\nu}_{\overline{\theta}}\}$ and $\{ \overline{\theta}^{\rm azi},\overline{\theta}^{\rm ele} \}$, and therefore estimating $\{\overline{\theta}^{\rm azi},\overline{\theta}^{\rm ele} \}$ is equivalent to estimating $\{ \overline{\mu}_{\overline{\theta}},\overline{\nu}_{\overline{\theta}} \}$. Taking $R_x(\alpha ) {\buildrel \Delta \over =} \left| \Xi_{\bar{N}_x}\left( \frac{2\pi d}{\lambda} \left( \alpha -\overline{\mu}_{\overline{\theta}} \right) \right) \right|$ in \eqref{CorFun} as an example, we report a tangible $R_x(\alpha )$ in Fig.~\ref{FigCorFun} and compare it with that of the CSA ($d = d_{\rm cri} = 0.5\lambda$). Accordingly, we summarize the following two properties of the WSA.

{\it a)~Angular Ambiguity.} Since $R_x( \alpha )$ attains its maximum value of $1$ when $\alpha = \overline{\mu}_{\overline{\theta}}$, it is intuitive to obtain an estimation of $\overline{\mu} \in ( -1,\,  1 )$ by determining an $\alpha$ within the range of $( -1, \, 1 )$ for maximizing $R_x( \alpha )$. However, this approach is unable to acquire the actual angle $\overline{\mu}_{\overline{\theta}}$ for WSA. This is because $R_x( \alpha )$ is periodic with the period $\frac{\lambda}{d}$, i.e., $R_x( \alpha ) = R_x\left( \alpha + \frac{\lambda}{d} \right)$, $\forall \alpha$. Therefore, the points within $\alpha \in ( -1,\, 1 )$ which maximize $R_x( \alpha )$ can be collected as
\begin{align}\label{AngAmb} 
{\cal{I}}_{\max,\overline{\mu}} {\buildrel \Delta \over =}& \left\{ \alpha \left| \arg \max\limits_{\alpha \in ( -1, \,1 )} R_x( \alpha ) \right. \right\}  
 = \left\{ \overline{\mu}_{\overline{\theta}} + \frac{k\lambda}{d}\left| k \in \mathbb{Z} \text{ and } -1 < \overline{\mu}_{\overline{\theta}} + \frac{k\lambda}{d} < 1 \right. \right\} .
\end{align}
For WSA with $d > 0.5\lambda$, there always exist some values of $\overline{\mu}_{\overline{\theta}}$ that yield ${\rm card}\big({\cal{I}}_{\max,\overline{\mu}}\big) > 1$. Worse still, when $d > \lambda$, ${\rm card}\big({\cal{I}}_{\max,\overline{\mu}}\big) > 1$ for all $\overline{\mu}_{\overline{\theta}} \in ( -1, \, 1 )$. This implies that we cannot determine the exact azimuth angle $\overline{\mu}_{\overline{\theta}}$ from the multi-element set ${\cal{I}}_{\max,\overline{\mu}}$. A similar discussion can be made for the elevation angle $\overline{\nu}_{\overline{\theta}}$. Therefore, an ambiguity inherently exists in determining $\{ \overline{\theta}^{\rm azi},\overline{\theta}^{\rm ele} \}$, which will cause excessive missed detections or false alarms for radar sensing.

{\it b)~Higher Angular Resolution.} Let us consider one period of $R_x( \alpha )$ without angular ambiguity, e.g., the region of $0 \le \alpha < \frac{2}{3}$ in Fig.~\ref{FigCorFun}. By letting $R_x( \alpha ) = 0$, we have
\begin{align}\label{HiAnRe} 
\alpha =& \overline{\mu}_{\overline{\theta}} + \frac{k\lambda}{\overline{N}_x d} , ~ k \in \mathbb{Z} , ~  {\rm mod}(k, \overline{N}_x) \neq 0 .
\end{align}
It indicates that the angular resolution of the WSA is $\frac{\lambda}{\overline{N}_x d}$. In other words, any two targets with different $\overline{\mu}_{\overline{\theta}}$ and $\overline{\mu}_{\overline{\theta}}'$ satisfying $\left| \overline{\mu}_{\overline{\theta}} - \overline{\mu}_{\overline{\theta}}' \right| \ge \frac{\lambda}{\overline{N}_x d}$ can be well separated by the WSA, as the mainlobes of their correlation functions would not influence each other. It can also be seen that as long as $\overline{N}_x d > N_x d_{\rm cri}$, a better angular resolution can be achieved by the WSA over the CSA even with a moderate $\overline{N}_x$. Moreover, it can be observed from Fig.~\ref{FigCorFun} that in the non-ambiguity region, the sidelobe amplitudes of $R_x( \alpha )$ associated with the WSA are much smaller than the corresponding sidelobe associated with the CSA. Similarly, the higher angular resolution of the WSA in elevation angle also inherently exists. This property of the WSA facilitates the CS-based algorithms for sparse signal recovery, since it suppresses the {\it power leakage} phenomenon \cite{MyTVT} and thus the associated angular-domain channels will be sparser in the non-ambiguity region.

It can be seen that the application of WSA is beneficial for achieving better angular resolution, provided that its inherent angular ambiguity is eliminated. We derive an improved CS-based channel reconstruction algorithm, which is tailored for the WSA to eliminate the angular ambiguity. To improve the sparse CE performance, one viable approach is to quantize the angular domain with different levels, and the task becomes identifying which sample is the closest to the real channel angle. These predefined levels form the so-called {\it dictionary} \cite{MyTVT}, which transforms the spatial domain to the angular domain. Taking the azimuth direction as an example, we design the samples in the angular domain as
\begin{align}\label{WSASamples} 
\overline{\psi}_g^{\rm{azi}} =& -1 + \frac{g \lambda}{\overline{G}_x d} , ~ g = 0,1,\cdots ,\overline{G}_x - 1 ,
\end{align}
which is an equally-spaced sampling within $\left[-1,\, -1+\frac{\lambda}{d}\right)$, where $\overline{G}_x \ge \overline{N}_x$ is the number of azimuth samples. Accordingly, the dictionary matrix of the WSA along the azimuth direction is given by
\begin{align}\label{WSADic} 
\overline{\bf{A}}_{\rm{azi}} =& \left[ \overline{\bf{a}}\left( \overline{\psi}_0^{\rm{azi}};\overline{N}_x \right) ~ \overline{\bf{a}}\left( \overline{\psi}_1^{\rm{azi}};\overline{N}_x \right) \cdots \overline{\bf{a}}\left( \overline{\psi}_{\overline{G}_x - 1}^{\rm{azi}};\overline{N}_x \right) \right] \in \mathbb{C}^{\overline{N}_x \times \overline{G}_x} .
\end{align}
The expression of $\overline{\bf{a}}\left( \overline{\psi}_g^{\rm{azi}};\overline{N}_x \right)$ is given in \eqref{WSAStrVecPart}. Note that $\overline{\bf{A}}_{\rm{azi}}$ will be a unitary matrix when $\overline{G}_x = \overline{N}_x$, which although guaranteeing the uniqueness of the angular-domain representation, suffers from limited resolution. To improve the sensing performance, we can set $\overline{G}_x > \overline{N}_x$ to increase the number of samples in the angular domain. With increased $\overline G_x$, the real channel angles are more likely to be close to some predefined samples in \eqref{WSASamples}, so that they can be estimated more accurately. We refer to $\overline{\bf{A}}_{\rm{azi}}$ as {\it redundant dictionary} when $\overline{G}_x > \overline{N}_x$ \cite{MyTVT}. Similarly, the dictionary matrix along the elevation direction can be obtained as ${\bf \overline{A}}_{\rm ele} \in \mathbb{C}^{\overline{N}_y \times \overline{G}_y}$, where $\overline{G}_y \ge \overline{N}_y$ is the number of elevation samples. The overall dictionary matrix for the WSA is obtained as $\overline{\bf{A}} = \overline{\bf{A}}_{\rm azi} \otimes \overline{\bf{A}}_{\rm ele} \in \mathbb{C}^{\overline{N} \times \overline{G_x} \overline{G}_y}$. Furthermore, the dictionary design in \eqref{WSASamples} and \eqref{WSADic} can be directly applied to the CSA by considering $d = d_{\rm cri} = 0.5\lambda$. Let $G_x^{\rm CU}$ ($G_x^{\rm UT}$) and $G_y^{\rm CU}$ ($G_y^{\rm UT}$) be the azimuth dimension and elevation dimension of the angular-domain channels, respectively, associated with the CU (UT). We can design the dictionary matrices ${\bf A}_{\rm CU} \in \mathbb{C}^{N \times G_x^{\rm CU}G_y^{\rm CU}}$ and ${\bf A}_{\rm UT} \in \mathbb{C}^{M \times G_x^{\rm UT}G_y^{\rm UT}}$ for the CU and the UT, respectively.

\begin{figure}[t]
\vspace*{-2mm}
\captionsetup{font={footnotesize,color={black}}, name = {Fig.}, singlelinecheck=off, labelsep = period}
\begin{center}
\includegraphics[width=4in]{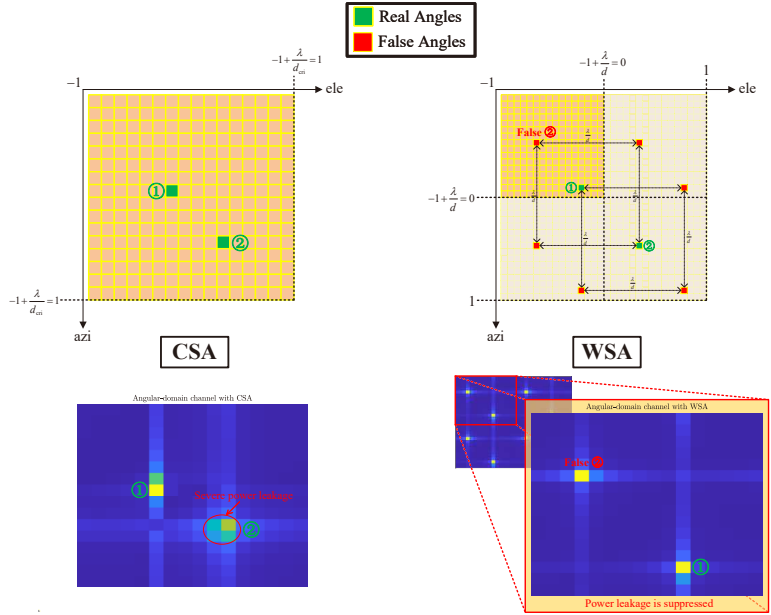}
\end{center}
\vspace*{-6mm}
\caption{The dictionary design and the associated angular-domain channels for radar sensing. We consider $2$ point targets marked by \ding{172} and \ding{173} respectively in the figure. $\overline{G}_x = \overline{G}_y =\overline{N}_x = \overline{N}_y = 16$, $d = d_{\rm cri}$ for the CSA, and $d = 2d_{\rm cri} = \lambda$ for the WSA.
Each real angle $\{\mu_{\rm real},\nu_{\rm real}\}$ and its corresponding false angle $\{\mu_{\rm false},\nu_{\rm false}\}$ satisfy ${\mu _{{\rm{false}}}} = {\mu _{{\rm{real}}}} + \frac{{{k_1}\lambda }}{d}$ and ${\nu _{{\rm{false}}}} = {\nu _{{\rm{real}}}} + \frac{{{k_2}\lambda }}{d}$, where $k_1,k_2 \in \mathbb{Z}$ so that ${\mu _{{\rm{false}}}},{\nu _{{\rm{false}}}} \in [-1, \, -1)$.
It is observed that by using the WSA, the power leakage phenomenon is suppressed, and thus the angular-domain channel is sparser, at the cost of angular ambiguity.}
\label{FigDiff} 
\vspace{-7mm}
\end{figure}

Although the dictionary designs for the WSA and CSA are semblable, it is necessary to highlight their differences, as illustrated in Fig.~\ref{FigDiff}. Unlike the dictionary for the CSA which covers the whole angular range $\big[-1,\, -1+\frac{\lambda}{d_{\rm cri}}\big)$, i.e., $\big[-1,1\big)$, the dictionary for the WSA in \eqref{WSASamples} only covers a smaller angular range $\big[-1,\, -1+\frac{\lambda}{d}\big)$, $d > \lambda/2$, to guarantee only one peak for each channel angle component in the angular domain at the cost of angular ambiguity.
Also observe from Fig.~\ref{FigDiff} that the angular-domain channel is likely to be sparser by using the WSA with higher angular resolution, as detailed in previous text.

Based on the designed dictionary matrices, the CIRs $\overline{\bf{H}}_l$ and ${\bf H}_l$, $0 \le l < L$, can be respectively represented as \cite{MyTVT,HeathJSAC}
\begin{align} 
\overline{\bf{H}}_l \approx & \overline{\bf{A}} \overline{\bf{H}}_l^{\rm A} {\bf A}_{\rm CU}^{\rm H} , \label{AdTrans1} \\
{\bf H}_l \approx & {\bf A}_{\rm UT} {\bf H}_l^{\rm A} {\bf A}_{\rm CU}^{\rm H} , \label{AdTrans2}
\end{align}
where $\overline{\bf{H}}_l^{\rm A} \in \mathbb{C}^{{\overline{G}_x}{\overline{G}_y} \times G_x^{\rm CU}G_y^{\rm CU}}$ and ${\bf H}_l^{\rm A} \in \mathbb{C}^{{{G}^{\rm UT}_x}{{G}^{\rm UT}_y} \times G_x^{\rm CU}G_y^{\rm CU}}$ are the angular-domain channels. By substituting \eqref{AdTrans1} and \eqref{AdTrans2} into \eqref{RadCSpro} and \eqref{CommCSFinal} respectively, we have
\begin{align} 
\label{Pro1}
\overline{\bf{Y}} \approx& \textsf{Q} \left\{ \overline{\bf{A}} \overline{\bf{H}}_{\rm AD} \left( {\bf I}_L \otimes {\bf A}_{\rm CU}^{\rm H} \right) \overline{\bf{\Phi}} + \overline{\bf{N}} \right\} , \\
\label{Pro2}
{\bf y}_{\rm valid} \approx& {\bf \Phi}_{\rm valid} \left( \left( {\bf I}_L \otimes {\bf A}_{\rm CU}^* \right) \otimes {\bf A}_{\rm UT} \right) {\rm vec}\big({\bf H}_{\rm AD}\big) + {\bf n}_{\rm valid} ,
\end{align}
where {$\overline{\bf{H}}_{\rm{AD}} = \left[ \overline{\bf{H}}_0^{\rm{A}} ~ \overline{\bf{H}}_1^{\rm{A}} \cdots \overline{\bf{H}}_{L - 1}^{\rm{A}} \right]$} and {${\bf{H}}_{\rm{AD}} = \left[ {\bf{H}}_0^{\rm{A}} ~ {\bf{H}}_1^{\rm{A}} \cdots {\bf{H}}_{L - 1}^{\rm{A}} \right]$} are the angular-delay {(abbreviated as the subscript ``AD'')} domain channels to be estimated, and they are sparse due to the well-known dual sparsity in both the angular domain and delay domain \cite{GaoCL,korean,XMa,HeathJSAC,XLin,MyTVT}.

\subsection{The Proposed CS-Based Algorithms}\label{S4.2}

\begin{algorithm}[!t]
\caption{Orthogonal Matching Pursuit with Support Refinement (OMP-SR)}
\label{ALG1}
{\bf Input}:
Echo pilot signals $\overline{\bf{Y}}$ (after quantized by low-resolution ADCs), measurement matrix $\overline{\bf{\Phi}}$, dictionary matrices $\overline{\bf{A}}$ and ${\bf A}_{\rm CU}$, and {\it stop criterion}.
\begin{algorithmic}[1]
\State {\bf Initialization}: ${\bf{\Psi}} = \left[ \overline{\bf{\Phi}}^{\rm T} \left( {\bf I}_L \otimes {\bf A}_{\rm CU}^{*} \right) \right] \otimes \overline{\bf{A}}$, ${\bf r} = {\rm vec}\big(\overline{\bf{Y}}\big)$, and ${\cal I} = {\cal I}_{\rm d} = {\cal I}_{\rm azi} = {\cal I}_{\rm ele} = \emptyset$.
\While{{\it stop criterion} is not met,}
	\State $i_{\rm supp} = \arg \max\limits_i  \left| \left[ {\bf{\Psi}}^{\rm H} {\bf{r}} \right]_i \right|$;
	\State ${\cal I} = {\cal I} \cup \left\{i_{\rm supp}\right\}$;
	\Statex \ \ \ \ {\it \% Function} ``ind2sub'' {\it below defined in \eqref{ind2sub}.}
	\State Obtain $i_{\rm AoA}$ and $i_{\rm aux}$ via $\left[i_{\rm AoA},i_{\rm aux}\right] = \text{ind2sub}\left(\left[\overline{G}_x \overline{G}_y,L G_x^{\rm CU} G_y^{\rm CU} \right],i_{\rm supp}\right)$;
	\State Obtain $i_{\rm AoD}$ and $i_{\rm d}$ via $\left[i_{\rm AoD},i_{\rm d}\right] = \text{ind2sub}\left(\left[G_x^{\rm CU} G_y^{\rm CU},L\right],i_{\rm aux}\right)$;
	\State Obtain $i^{\rm azi}_{\rm AoD}$ and $i^{\rm ele}_{\rm AoD}$ via $\left[i^{\rm ele}_{\rm AoD},i^{\rm azi}_{\rm AoD}\right] = \text{ind2sub}\left(\left[G_y^{\rm CU}, G_x^{\rm CU}\right],i_{\rm AoD}\right)$;
	\State Obtain $i^{\rm azi}_{\rm AoA}$ and $i^{\rm ele}_{\rm AoA}$ via $\left[i^{\rm ele}_{\rm AoA},i^{\rm azi}_{\rm AoA}\right] = \text{ind2sub}\left(\left[\overline G_y, \overline G_x\right],i_{\rm AoA}\right)$;
	\State ${\cal I}_{\rm delay} = {\cal I}_{\rm delay} \cup \{ (i_{\rm d} - 1)T_{\rm s} - \tau_{\rm p}\}$; \ \ {\it \% Delay Estimation.}
	\Statex \ \ \ \ {\it \% Angle Estimation (steps 10--15):}
	\State $\widehat{\mu}_{\rm C} = -1+2\left(i^{\rm azi}_{\rm AoD}-1\right)/{G_x^{\rm CU}}$;
	\State {$\widehat{\mu}_{\rm F} = -1+\left(i^{\rm azi}_{\rm AoA}-1\right)\lambda /\overline{G}_x d$};
	\State {$\widehat{\mu}_{\rm F} = \widehat{\mu}_{\rm F} + \frac{k\lambda}{d}$}, where $k \in \mathbb{Z}$ is the integer closest to $\left(\widehat{\mu}_{\rm C} - \widehat{\mu}_{\rm F}\right) \frac{d}{\lambda}$;  \ \ {\it \% Spatial Consistency.}
	\State ${\cal I}_{\rm azi} = {\cal I}_{\rm azi} \cup \left\{ \widehat{\mu}_{\rm F}\right\}$;
	\State Obtain finer estimation of elevation angle $\widehat{\nu}_{\rm F}$ similarly to steps 10--12;
	\State ${\cal I}_{\rm ele} = {\cal I}_{\rm ele} \cup \left\{ \widehat{\nu}_{\rm F}\right\}$;
	\Statex \ \ \ \ {\it \% Support Refinement (steps 16--17):}
	\State ${\bf a}_{\rm aux} = {\bf a}\big(\widehat \mu_{\rm F};N_x\big) \otimes {\bf a}\big(\widehat \nu_{\rm F};N_y\big)$, ${\bf \Phi}_{\rm aux} = \left[{\bf \overline \Phi}^{\rm T}\right]_{\{[(i_{\rm d}-1)N+1]:i_{\rm d}N\}}$, and ${\bf \overline{a}}_{\rm aux} = {\bf \overline{A}}_{\{i_{\rm AoA}\}}$;
	\State Replace $i_{\rm supp}$-th column of $\bf{\Psi}$ by ${\bf q} = \big({\bf{\Phi}}_{\rm aux}{\bf a}_{\rm aux}^*\big) \otimes \overline{\bf{a}}_{\rm aux}$;
	\State $\widehat{\bf{g}} = \left[ {\bf{\Psi}} \right]_{\cal I}^\dag {\rm vec}\big(\overline{\bf{Y}}\big)$;
	\State ${\bf r} = {\rm vec}\big(\overline{\bf{Y}}\big) - \left[ {\bf{\Psi}} \right]_{\cal I} \widehat{\bf{g}}$;
\EndWhile
\State Reconstruct estimate $\widehat{\overline{\bf{H}}}_{\rm SD}$ of $\overline{\bf{H}}_{\rm SD}$ via \eqref{RadarCh} and \eqref{CIR} based on ${\cal I}_{\rm d}$, ${\cal I}_{\rm azi}$, ${\cal I}_{\rm ele}$, and $\widehat{\bf{g}}$;
\end{algorithmic}
{\bf Output}:
Estimate $\widehat{\overline{\bf{H}}}_{\rm SD}$ of $\overline{\bf{H}}_{\rm SD}$, estimated target delays ${\cal I}_{\rm d}$, estimated azimuth angles ${\cal I}_{\rm azi}$, and estimated elevation angles ${\cal I}_{\rm ele}$.
\end{algorithm}

We now detail our proposed CS-based algorithms for solving the radar sensing problem \eqref{Pro1} and the CE problem \eqref{Pro2}, respectively.

{\it a)~Radar Sensing Problem.} As mentioned previously, the radar CIR $\overline{\bf{H}}_{\rm SD}$, or equivalently $\overline{\bf{H}}_{\rm AD}$, is estimated at the DFRC station which has sufficient computing power.
We can formulate this radar sensing problem as a sparse signal recovery problem as follow:
\begin{align}\label{SparsePro} 
\begin{array}{rl}
\min\limits_{\overline{\bf{H}}_{\rm AD}} & \left\| \overline{\bf{Y}} - \overline{\bf{A}} \overline{\bf{H}}_{\rm{AD}} \left( {\bf{I}}_L \otimes {\bf{A}}_{\rm{CU}}^{\rm H} \right) \overline{\bf{\Phi}} \right\|_F^2 ,  \\
\text{s.t.} & \left\| \overline{\bf{H}}_{\rm AD} \right\|_0 < \varepsilon , \text{ and spatial consistency holds},
\end{array}
\end{align}
where $\varepsilon$ is the threshold defining the {\it stop criterion}. Our proposed OMP-SR algorithm for solving the optimization in \eqref{SparsePro} is summarized in \textbf{Algorithm~\ref{ALG1}}, which takes advantage of the spatial consistency and the mixed angular resolutions induced by the WSA and CSA.
For notational convenience, the function $\left[ I,J\right] = \text{ind2sub}\left(\left[X,Y\right],Z\right)$ in \textbf{Algorithm~\ref{ALG1}} is defined here:
\begin{align}\label{ind2sub} 
\begin{split}
I & = Z - (\left\lceil Z/X \right\rceil-1)X , \\
J & = \left\lceil Z/X \right\rceil ,
\end{split}
\end{align}
which helps to extract the indices of the azimuth angle, the elevation angle, and the delay-offset based on the selected atom (see, e.g., steps 5--8 of \textbf{Algorithm~\ref{ALG1}}).

The main differences between \textbf{Algorithm~\ref{ALG1}} and the traditional OMP algorithm \cite{korean} lie in the following two aspects:
(i)~By identifying the atom position with the most significant correlation (step 3), we obtain an estimation of delay, a coarse estimation of angle $\widehat{\mu}_{\rm C}$, and a finer estimation of angle $\widehat{\mu}_{\rm F}$ (but with ambiguity). Given the fact that $\widehat{\mu}_{\rm C}$ and $\widehat{\mu}_{\rm F}$ correspond to the same angle component due to the spatial consistency, we refine $\widehat{\mu}_{\rm F}$ by adding a term of $\frac{\lambda}{d}$ multiplied by an integer to it so that it approaches the coarse estimation $\widehat{\mu}_{\rm C}$, and finally obtain the finer estimation of angle without ambiguity (step 12).
(ii)~With the finer angle estimation, we reconstruct the steering vector at the CSA side (step 16) and replace the corresponding column in the sensing matrix $\bf{\Psi}$ by using this refined steering vector (step 17). The sensing matrix after refinement will better model the angular-domain channel. The refined sensing matrix $\bf{\Psi}$ is used for the following subspace project (step 18) and residual update (step 19). By setting an appropriate stop criterion, our OMP-SR algorithm will output the estimates of $\overline{\bf{H}}_{\rm SD}$ and the parameters of interest ${\cal I}_{\rm d}$, ${\cal I}_{\rm azi}$, and ${\cal I}_{\rm ele}$.
Note that for point target identification, the parameter estimates ${\cal I}_{\rm d}$, ${\cal I}_{\rm azi}$, and ${\cal I}_{\rm ele}$ are vital, while for radar imaging application, the estimate of $\overline{\bf{H}}_{\rm SD}$ can provide more information, e.g., the shapes and types of the targets of interest. 

As for the {\it stop criterion} of \textbf{Algorithm~\ref{ALG1}}, a widely-adopted method is comparing the energy of residual with a pre-defined threshold \cite{korean}. However, the optimal threshold of this residual-based criterion is hard to obtain, especially in the high-dynamic ISAC scenarios.
With an inappropriate threshold, the number of algorithm iterations may be either too large (with unaffordable computational burden) or too small (sufficient precision cannot be guaranteed). A practical alternative is setting a maximum number of iterations \cite{MyTcom}, which makes both the running time and performance of the algorithm predictable. In view of this, we will adopt the stop criterion in \cite{MyTcom}, where the maximum number of iterations will be experimentally obtained (see Fig. \ref{FigIter}).
	
{\it b)~CE Problem.}
Since the analog architecture is deployed, the dimension of the received pilot signals in each time slot is limited to $1$ at the UT, which decreases the number of measurements and makes it hard to recover the whole communication CIR. Moreover, the energy-constrained UT has limited computational capability compared with the DFRC station. Given the above two limitations, only the LoS angles, $\left\{ \theta^{\rm azi}_{\rm LoS},\theta^{\rm ele}_{\rm LoS}\right\}$, $\left\{ \phi^{\rm azi}_{\rm LoS},\phi^{\rm ele}_{\rm LoS}\right\}$, and the LoS delay $\tau_{\rm LoS}$ in ${\bf{H}}_{\rm{SD}}$ are estimated via \eqref{CommCSFinal} at the UT.
After performing the estimation, the UT needs to feed the estimate of $\left\{ \phi^{\rm azi}_{\rm LoS},\phi^{\rm ele}_{\rm LoS}\right\}$ back to the CU. The UT and CU then conduct beamforming based on the estimated $\left\{ \theta^{\rm azi}_{\rm LoS},\theta^{\rm ele}_{\rm LoS}\right\}$ and $\left\{ \phi^{\rm azi}_{\rm LoS},\phi^{\rm ele}_{\rm LoS}\right\}$, respectively, to guarantee reliable data transmission.

\begin{algorithm}[!t]
\caption{Low-complexity Channel Estimation at UT}
\label{ALG2}
{\bf Input}:
Receive pilot signals ${\bf{y}}_{\rm valid}$, measurement matrix ${\bf{\Phi}}_{\rm valid}$, dictionaries ${\bf{A}}_{\rm UT}$ and ${\bf A}_{\rm CU}$.
\begin{algorithmic}[1]
\State $i_{\rm supp} = \arg \max\limits_i  \left| \left[  \left( \left( {\bf I}_L \otimes {\bf A}_{\rm CU}^{\rm T} \right) \otimes {\bf A}^{\rm H}_{\rm UT} \right) {\bf{\Phi}}^{\rm H}_{\rm valid} {\bf{y}}_{\rm valid} \right]_i \right|$;
\Statex{\it \% Function} ``ind2sub'' {\it below defined in \eqref{ind2sub}.}
\State Obtain $i_{\rm UT}$ and $i_{\rm aux}$ via $\left[i_{\rm UT},i_{\rm aux}\right] = \text{ind2sub}\left(\left[ G^{\rm UT}_x G^{\rm UT}_y,L G_x^{\rm CU} G_y^{\rm CU}\right],i_{\rm supp}\right)$;
\State Obtain $i_{\rm CU}$ and $i_{\rm d}$ via $\left[i_{\rm AoD},i_{\rm d}\right] = \text{ind2sub}\left(\left[G_x^{\rm CU} G_y^{\rm CU},L\right],i_{\rm aux}\right)$;
\State Obtain $i^{\rm azi}_{\rm CU}$ and $i^{\rm ele}_{\rm CU}$ via $\left[i^{\rm ele}_{\rm CU},i^{\rm azi}_{\rm CU}\right] = \text{ind2sub}\left(\left[G_y^{\rm CU}, G_x^{\rm CU}\right],i_{\rm CU}\right)$;
\State Obtain $i^{\rm azi}_{\rm UT}$ and $i^{\rm ele}_{\rm UT}$ via $\left[i^{\rm ele}_{\rm UT},i^{\rm azi}_{\rm UT}\right] = \text{ind2sub}\left(\left[G^{\rm UT}_y, G^{\rm UT}_x\right],i_{\rm UT}\right)$;
\State {$\widehat{\mu}_{\rm UT} = -1+2\big(i^{\rm azi}_{\rm UT}-1\big)/G_x^{\rm UT}$}, {$\widehat{\nu}_{\rm UT} = -1+2\big(i^{\rm ele}_{\rm UT}-1\big)/G_y^{\rm UT}$}, {$\widehat{\mu}_{\rm CU} = -1+2\big(i^{\rm azi}_{\rm CU}-1)\big/G_x^{\rm CU}$},	and {$\widehat{\nu}_{\rm CU} = -1+2\big(i^{\rm ele}_{\rm CU}-1\big)/G_y^{\rm CU}$}; \ \ {\it \% Angle Estimation.}
\State $\widehat{\tau} = \left\{ \big(i_{\rm d} - 1\big)T_{\rm s} - \tau_{\rm p}\right\}$;\ \ {\it \% Delay Estimation.}
\end{algorithmic}
{\bf Output}: Estimates of LoS (virtual) angles $\left\{ \widehat{\mu}_{\rm UT}, \widehat{\nu}_{\rm UT}\right\}$ at UT side and $\left\{ \widehat{\mu}_{\rm CU}, \widehat{\nu}_{\rm CU}\right\}$ at CU side, and estimate of LoS delay-offset $\widehat{\tau}$.
\end{algorithm}

The procedure of the CE is summarized in \textbf{Algorithm~\ref{ALG2}}, which is a single correlation step of the OMP framework, and it returns the position of the most significant atom (i.e., the LoS path component) in $\bf{H}_{\rm AD}$. This single step solution has low computational complexity and makes it practical for implementation at the UT. Note that one can readily extend \textbf{Algorithm~\ref{ALG2}} to recover the whole channel ${\bf H}_{\rm AD}$ for more sophisticated beamforming methods. However, it is more practical for the energy-constrained UT to conduct the beamforming based on the LoS angles obtained by the low-complexity CE methods such as  \textbf{Algorithm~\ref{ALG2}}.

\subsection{Doppler Estimation Framework}\label{S4.3}

The Doppler estimation and compensation are essential for effective communication-centric ISAC systems. We focus on the Doppler estimation for uplink communication, and the proposed method can be directly applied to the speed estimation of radar targets.
With the estimates obtained at the previous stage, the DFRC station will first conduct a UT scheduling, i.e., choose at most $N_{\rm RF}$ UTs which are well-separated in either the angular-domain or the delay-domain to serve, in order to avoid the severe inter-user interference (IUI).
In the rest of this subsection, the index $u$, $1\le u \le U$, is introduced to represent the $u$-th UT out of $U$ scheduled UTs. Without loss of generality, we assume the number of served UTs to be that of RFCs at the CU, i.e., $U = N_{\rm RF}$. We assume that only the channel coefficients $\left\{ g_{\rm{LoS}}(t),g_{c,l}(t),\overline{g}_{c,l}(t) \right\}$ vary over different pilot signals for Doppler estimation (see the discussion in Subsection~\ref{S3.1}). During the payload data transmission, the analog beamformer of the $u$-th UT ${\bf w}_{{\rm D},u} \in \mathbb{C}^{M \times 1}$ is given as
\begin{align}\label{temp1} 
{\bf w}_{{\rm D},u} =& {\bf a}\left(\widehat{\mu}_{{\rm UT},u},M_x\right) \otimes {\bf a}\left(\widehat{\nu}_{{\rm UT},u},M_y\right) ,
\end{align}
which is a commonly-used beam steering scheme \cite{TcomSurvey,FLiuTWC} to obtain the highest beamforming gain at certain direction, while the hybrid beamformer at the CU ${\bf F}_{\rm D} \in \mathbb{C}^{N \times N_{\rm RF}}$ is given by
\begin{align}\label{temp2} 
{\bf F}_{{\rm D}} =& \left[{\bf f}_{{\rm D},1} \cdots {\bf f}_{{\rm D},U}\right] ,
\end{align}
where ${\bf f}_{{\rm D},u} = {\bf a}\left(\widehat{\mu}_{{\rm CU},u},N_x\right) \otimes {\bf a}\left(\widehat{\nu}_{{\rm CU},u},N_y\right)$. 

\begin{figure}[t]
\vspace*{-3mm}
\captionsetup{font={footnotesize}, name = {Fig.}, singlelinecheck=off, labelsep = period}
\begin{center}
\includegraphics[width=4in]{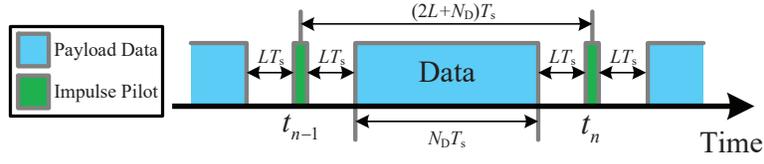}
\end{center}
\vspace*{-7mm}
\caption{The detailed payload data transmission frame structure in Fig.~\ref{FigFrmStru}.}
\label{FigDetail} 
\vspace{-7mm}
\end{figure}

As illustrated in Fig.~\ref{FigFrmStru}, a small amount of pilot signals are inserted between the data frames to estimate the Doppler frequencies. These pilot signals are depicted in Fig.~\ref{FigDetail}. Each UT emits an uplink impulse pilot signal (marked by green pulse in Fig.~\ref{FigDetail}) between every two data frames, and each data frame has the duration $N_{\rm D}T_{\rm s}$. The zero-padding of duration $LT_{\rm s}$ is added on both sides of the impulse pilot signal to eliminate the inter-frame interference. With this arrangement, the received $n$-th pilot signal ${\bf y}^{\rm D}(\tau, t_n) \in \mathbb{C}^{N_{\rm RF} \times 1}$ is expressed as
\begin{align}\label{y} 
{\bf y}^{\rm D}(\tau ,t_n) =& {\bf F}^{\rm T}_{\rm D} \sum\limits_{u = 1}^U \sqrt{P^{\rm UT}_u} {\bf{H}}_u^{\rm T}\left(\tau ,t_n \right) {\bf{w}}_{{\rm{D}},u}^{*} + {\bf F}^{\rm T}_{\rm D} {\bf n}^{\rm D}(\tau ,t_n) ,
\end{align}
where $t_n$ is the time index of the $n$-th impulse, ${\bf H}^{\rm T}_u(\tau,t_n) \in \mathbb{C}^{N \times M}$ is the uplink communication channel of the $u$-th UT (the channel reciprocal in TDD mode is considered), ${P}^{{\rm UT}}_u$ is the transmit power of the $u$-th UT, and ${\bf n}^{\rm D}(\tau,t_n) \sim {\cal CN}\big({\bf 0}_{N \times 1},\sigma_{\rm n}^2 {\bf I}_N\big)$ is the AWGN vector.

\begin{remark}\label{REM3}
Due to the high path loss of NLoS paths in the mmWave band, the NLoS paths contribute little for the communication applications in the presence of the strong LoS path. For example, the Rician factor is considered as 20\,dB in the mmWave band \cite{GaoCL}. Moreover, the beam steering towards the LoS direction in \eqref{temp1} and \eqref{temp2} will significantly suppress the signals from the NLoS paths, benefiting from the asymptotic orthogonality of massive MIMO \cite[Lemma 1]{FLiuTWC}.
\end{remark}

According to \textbf{Remark~\ref{REM3}}, we can safely treat the influence of the NLoS paths as some negligible noises after beamforming. Hence, we can re-write \eqref{y} as
\begin{align}\label{yLoS} 
{\bf y}^{\rm D}(\tau ,t_n) \approx & \sum\limits_{u = 1}^U \sqrt{P^{\rm{UT}}_u} \underbrace{ {\bf{F}}_{\rm{D}}^{\rm T} {\bf{a}}_N^{*}\left( \varphi_{{\rm{LoS}},u}^{\rm{azi}},\varphi_{{\rm{LoS}},u}^{\rm{ele}} \right) }_{{\bf{g}}_{{\rm LoS},u}^{\rm{CU}}} \underbrace{ {\bf{a}}_M^{\rm T}\left( \theta_{{\rm{LoS}},u}^{\rm{azi}},\theta_{{\rm{LoS}},u}^{\rm{ele}} \right) {\bf{w}}_{{\rm{D}},u}^{*} }_{G_{u}^{\rm{UT}}} \nonumber \\ 
& \times  p\left( \tau - \tau_{{\rm{LoS}},u} - \tau_{\rm{p}} \right) g_{{\rm{LoS}},u}\left( t_n \right) + {\bf F}^{\rm T}_{\rm D} {\bf n}^{\rm D}(\tau,t_n) \nonumber \\
=& \sum\limits_{u = 1}^U \sqrt{P^{\rm{UT}}_u} G_u^{\rm{UT}} {\bf{g}}_{{\rm{LoS}},u}^{\rm{CU}} p\left( \tau - \tau_{{\rm{LoS}},u} - \tau_{\rm{p}} \right) g_{{\rm{LoS}},u}\left( t_n \right) + {\bf{F}}_{\rm{D}}^{\rm T} {\bf{n}}^{\rm{D}}\left(\tau ,t_n\right) .
\end{align}
where $G_u^{\rm{UT}}$ is the transmit beamforming gain at the $u$-th UT, and
\begin{align}\label{yLOSg} 
{\bf{g}}_{{\rm{LoS}},u}^{\rm{CU}} =& \left[ {\bf{a}}_N^{\rm H}\left( \varphi_{{\rm{LoS}},u}^{\rm{azi}},\varphi_{{\rm{LoS}},u}^{\rm{ele}} \right) {\bf{f}}_{{\rm{D}},1} \cdots {\bf{a}}_N^{\rm H}\left( \varphi_{{\rm{LoS}},u}^{\rm{azi}},\varphi_{{\rm{LoS}},u}^{\rm{ele}} \right) {\bf{f}}_{{\rm{D}},U} \right]^{\rm T} .
\end{align}
Note that only the LoS components in ${\bf H}^{\rm T}_u(\tau,t_n)$ are kept in \eqref{yLoS}. Furthermore, by utilizing the estimated LoS delay-offset $\widehat{\tau}_u$ of the $u$-th UT's channel from \textbf{Algorithm~\ref{ALG2}}, we obtain the received pilot signal in the delay domain which corresponds to the LoS path as
\begin{align}\label{yDelay} 
\left[ {\bf y}^{\rm D}\left(\widehat{\tau}_u ,t_n\right) \right]_u =& \sqrt{P^{\rm{UT}}_u} G_u^{\rm{UT}} \left[{\bf{g}}_{{\rm LoS},u}^{\rm{CU}}\right]_u p\left( \widehat{\tau}_u - \tau_{{\rm{LoS}},u} - \tau_{\rm{p}} \right) g_{{\rm{LoS}},u}\left( t_n \right) \nonumber \\
& \hspace*{-10mm}+ \underbrace{ \sum\limits_{u' \ne u} \sqrt{P^{\rm{UT}}_{u'}} G_{u'}^{\rm{UT}} \left[ {\bf{g}}_{{\rm LoS},u'}^{\rm{CU}} \right]_u p\left( \widehat{\tau}_u - \tau_{{\rm{LoS}},u'} - \tau_{\rm{p}} \right) g_{{\rm{LoS}},u'}\left( t_n \right) }_{\text{IUI term}} + {\bf{f}}_{\rm{D}}^{\rm T} {\bf{n}}^{\rm{D}}\left(\widehat{\tau}_u,t_n\right) .
\end{align}
It can be seen that in \eqref{yDelay} the IUI is significantly suppressed. This is because given the well-separated UTs after scheduling, we usually can guarantee 
\begin{align}\label{ADOrtho} 
\left[ {\bf{g}}_{{\rm{LoS}},u'}^{\rm{CU}} \right]_u \approx & 0 \text{ and/or } p\left( \widehat{\tau}_u - \tau_{{\rm{LoS}},u'} - \tau_{\rm{p}} \right) \approx 0 , \text{ for } \forall u' \neq u .
\end{align}
The formula in \eqref{yDelay} motivates us to estimate the Doppler frequencies in a UT-wise manner, where the IUI can be treated as noise for each UT. By collecting $\left\{ \left[ {\bf{y}}^{\rm{D}}\left(\widehat{\tau}_u,t_n\right)\right]_u \right\}_{n=1}^{P_{\rm D}}$ in $P_{\rm D}$ successive impulse pilot signals, we obtain a time series ${\bf y}^{\rm D}_u \in \mathbb{C}^{P_{\rm D}\times 1}$ as
\begin{align}\label{TS} 
{\bf y}^{\rm D}_u  =& \left[ \left[ {\bf{y}}^{\rm{D}}\left(\widehat{\tau}_u,t_1\right) \right]_u \cdots \left[ {\bf{y}}^{\rm{D}}\left(\widehat{\tau}_u,t_{P_{\rm D}}\right) \right]_u \right]^{\rm T} 
 = A_u \left[ g_{\rm{LoS},u}\left( t_1 \right) \cdots g_{\rm{LoS},u}\left( t_{P_{\rm D}} \right) \right]^{\rm T}  + {\bf n}^{\rm eff}_u ,
\end{align}
where $A_u = \sqrt{P^{\rm{UT}}_u} G_u^{\rm{UT}}\left[{\bf{g}}_{\rm{LoS},u}^{\rm{CU}}\right]_u p\left( \widehat{\tau}_u  -\tau_{\rm{LoS},u} - \tau_{\rm{p}} \right)$ is a constant, ${\bf n}_u^{\rm eff} \in \mathbb{C}^{P_{\rm D}\times 1}$ is an effective noise vector including both the IUI and AWGN in \eqref{yDelay}, and $t_n - t_{n - 1} = \left( 2L + N_{\rm{D}} \right) T_{\rm{s}}$, $n>1$. The time series ${\bf y}^{\rm D}_u$ can be viewed as the noisy uniformly-spaced samples of a single-tone complex sinusoid, whose frequency is the Doppler frequency of the $u$-th UT. The sampling interval is $T_{\rm D} {\buildrel \Delta \over =} \left( 2L + N_{\rm{D}} \right) T_{\rm{s}}$. Many off-the-shelf techniques can be used for estimating the Doppler frequency via ${\bf y}^{\rm D}_u$. We resort to the weighted normalized auto-correlation linear predictor (WNALP) \cite{WNALP} for its near-optimal performance.

\begin{remark}\label{REM4}
The proposed Doppler estimation framework for multi-user uplink communications can be directly applied to the speed estimation of radar targets. In the context of radar, the CU and RU align the beams towards the targets of interest by using the estimation results at the initial stage, as done in \eqref{temp2}. During the target tracking, the CU emits the impulse pilot signal with an appropriate repetitive interval, and the RU receives the corresponding echo signals for the Doppler estimation, as done in \eqref{y}-\eqref{TS}. The Doppler estimation for communications and radar sensing can be conducted in a time-division manner to avoid cross interference between the uplink communication signals and radar echo signals.
\end{remark}

\subsection{Computational Complexity Analysis}

In this subsection, we analyse the computational complexity of the proposed ISAC scheme as follows. 
\begin{itemize}
{\item OMP-SR in \textbf{Algorithm~\ref{ALG1}} has four major parts: correlation (step 3), support refinement (steps 16--17), project subspace (step 18), and residual update (step 19), and the computational complexity of each part is $\mathcal{O}\left(Q\overline{N}L\overline{G}_x \overline{G}_y G^{\rm CU}_x G^{\rm CU}_y \right)$, $\mathcal{O}\left(Q\overline{N}+(Q+1)N\right)$, $\mathcal{O}\left(I^3+2Q\overline{N}I^2+Q\overline{N}I\right)$, and $\mathcal{O}\left(Q\overline{N}I\right)$, respectively, where $I$ stands for the current number of iterations.}

{\item Low-complexity CE scheme in \textbf{Algorithm~\ref{ALG2}} has the overall computational complexity of $\mathcal{O}\left(N_{\rm valid}G^{\rm CU}_x G^{\rm CU}_y G^{\rm UT}_x G^{\rm UT}_y\right)$, where $N_{\rm valid}$ is the dimension of ${\bf y}_{\rm valid}$.}

{\item The proposed Doppler estimation framework has the computational complexity of $\mathcal{O}\left(UP_{\rm D}\right)$.}
\end{itemize}

\section{Simulation Results}\label{S5}

In this section, we present numerical results to evaluate the performance of the proposed ISAC scheme, and compare it with existing counterparts in the literature.

\subsection{Experimental Setting}\label{S5.1}

We consider a vehicular network with a DFRC station as the RSU \cite{FLiuTWC}. Since in this case, the system only needs to acquire the information of horizontal obstacles to avoid collision, we focus only on the azimuth angles of UTs and targets by setting $N_y\! =\! \overline{N}_y\! =\! 1$. Note that our ISAC scheme is also valid for the full-dimensional CE and radar sensing. In our simulation system, we set $N_x\! =\! 16$, $N_{\rm RF}\! =\! 4$, $\overline{N}_x\! =\! M_x\! =\! 8$, $f_{\rm c}\! =\! 77$\,GHz, $T_{\rm s}\! =\! 5\times 10^{-9}$\,s (bandwidth ${\rm BW}\! =\! 200$\,MHz), $L\! =\! 32$, and $T_{\rm GI}\! =\! 10$ \cite{HeathJSAC}. Each ADC at the RU uniformly quantizes the receive signals to $2^B$ levels with $B$ quantization bits. The raised cosine filter with a roll-off factor of $0.8$ and single side duration $\tau_{\rm p}\! =\! 6 T_{\rm s}$ is adopted as $p(\tau )$. The noise-power spectral density at the receivers is $\text{NPSD}\! =\! -174$\,dBm/Hz, and the power of AWGN $\sigma_{\rm n}^2$ is thus $\sigma_{\rm n}^2\! =\! \text{NPSD} \times {\rm BW}\! \approx\! -91$\,dBm. The number of sub-path components in each clustered target or scatter is $\overline{N}_{\rm P} = N_{\rm P} = 15$, and the central azimuth angle, central elevation angle, and central delay-offset of each cluster follow ${\cal U}[0,\,2\pi)$, ${\cal U}[0,\,\pi/3]$, and ${\cal U}[0,\,(L\! -\! 1)T_{\rm s}\! -\! 2\tau_{\rm p}]$, respectively. Each cluster is generated with an angle spread $7.5^\circ$ and a delay spread $0.3T_{\rm s}$. The time-varying channel coefficients $g_{\rm LoS}(t)$, $g_{c,l}(t)$, and $\overline{g}_{c,l}(t)$ are given as
\begin{align} 
\label{temp3}
g_{\rm LoS}(t) & = \frac{\lambda e^{\textsf{j}\theta}}{4\pi d_{\rm{UT}}} e^{\textsf{j}2\pi f_{\rm D} t} , \\
\label{temp4}
g_{c,l}(t) & = \frac{\left| g_{\rm{LoS}}(t) \right| e^{\textsf{j}\theta_{c,l}}}{\sqrt{K_{\rm{f}} N_{\rm{C}} N_{\rm{P}}}} e^{\textsf{j}2\pi f_{{\rm D},c} t} , \\
\label{temp5}
\overline{g}_{c,l}(t) & = \sqrt{\frac{\overline{\sigma}_c \lambda^2}{N_{\rm{P}} (4\pi )^3 \overline{d}_c^4}} e^{\textsf{j}\overline{\theta}_{c,l}} e^{\textsf{j} 2\pi \overline{f}_{{\rm D},c} t} ,
\end{align}
where $d_{\rm UT}$ ($\overline{d}_c$) is the distance between the UT (the $c$-th target) and the DFRC station, $\theta$, $\theta_{c,l}$, and $\overline{\theta}_{c,l}$ are the phase-shifts induced by the corresponding channel paths, while $f_{\rm D}$, $f_{{\rm D},c}$, and $\overline{f}_{{\rm D},c}$ are the Doppler frequencies of the UT, the $c$-th scatter, and the $c$-th target, respectively. $K_{\rm f}$ is the Rician factor, and $\overline{\sigma}_c$ is the RCS of the $c$-th target. We set $d_{\rm UT}\! \sim\! {\cal U}[10,\, 20]$\,m, $\overline{d}_c\! \sim\! {\cal U}[5,\,10]$\,m, $\theta, \theta_{c,l}, \overline{\theta}_{c,l}\! \sim\! {\cal U}[0,\,2\pi)$, $f_{\rm D}, f_{{\rm D},c}, \overline{f}_{{\rm D},c}\! \sim\! {\cal U}[-7.1,\,7.1]$\,kHz (corresponding to a maximum radical velocity $100$\,km/h), $K_{\rm f}\! =\! 20$\,dB, and $\overline{\sigma}_c\! \sim\! {\cal U}[0.5,\, 5]$\,$\text{m}^2$. The length of each payload data frame is $N_{\rm D}\! =\! 1024$ (see Fig.~\ref{FigDetail}), and without loss of generality we set $t_1\! =\! 0$ in \eqref{TS} for the Doppler estimation. {Unless stated otherwise}, $P\! =\! 200$, $P_{\rm DL}\! =\! 60$\,dBm, $\overline{N}_{\rm C}\! =\! N_{\rm C}\! =\! 6$, $d\! =\! 1.5\lambda$, $\overline{G}_x / \overline{N}_x\! =\! 2$, $B\! =\! 5$, $T^{\rm CU}_{\rm RF}\! =\! T^{\rm UT}_{\rm RF}\! =\! 30$, and {the number of iterations is set to $150$ in \textbf{Algorithm~\ref{ALG1}}.}

\subsection{Numerical Results}\label{S5.2}

\subsubsection{Radar sensing performance}\label{S5.2.1}

\begin{figure}[tp!]
\vspace*{-2mm}
\captionsetup{font={footnotesize}, name = {Fig.}, singlelinecheck=off, labelsep = period}
{\begin{minipage}[t]{0.49\linewidth}
\centering
\includegraphics[width=2.6in,height=2.05in]{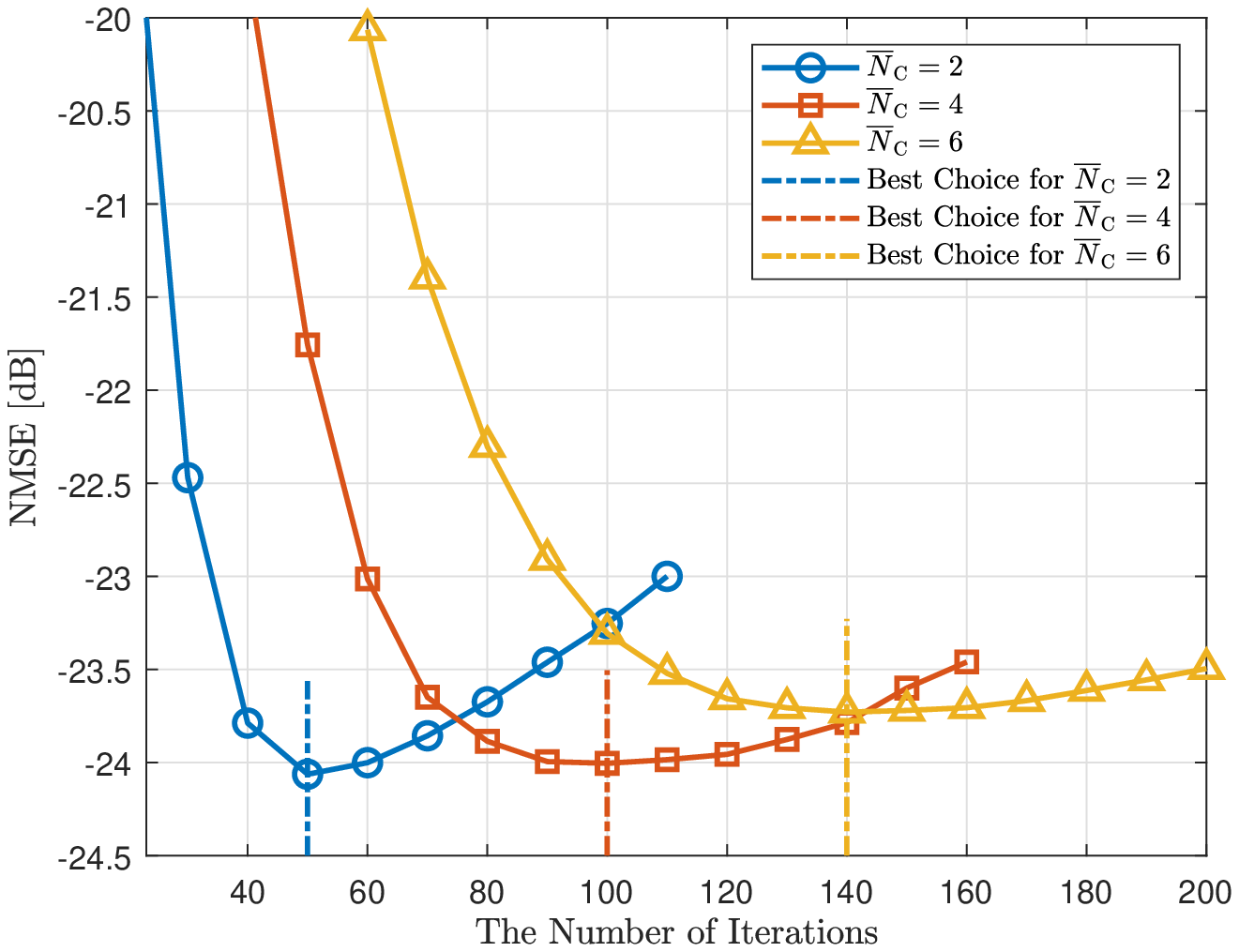}
\vspace*{-3mm}
\caption{The convergence of OMP-SR algorithm. Three different environmental conditions are compared.}
\label{FigIter} 
\end{minipage}}
\hfill
{\begin{minipage}[t]{0.49\linewidth}
\centering
\includegraphics[width=2.6in,height=2.05in]{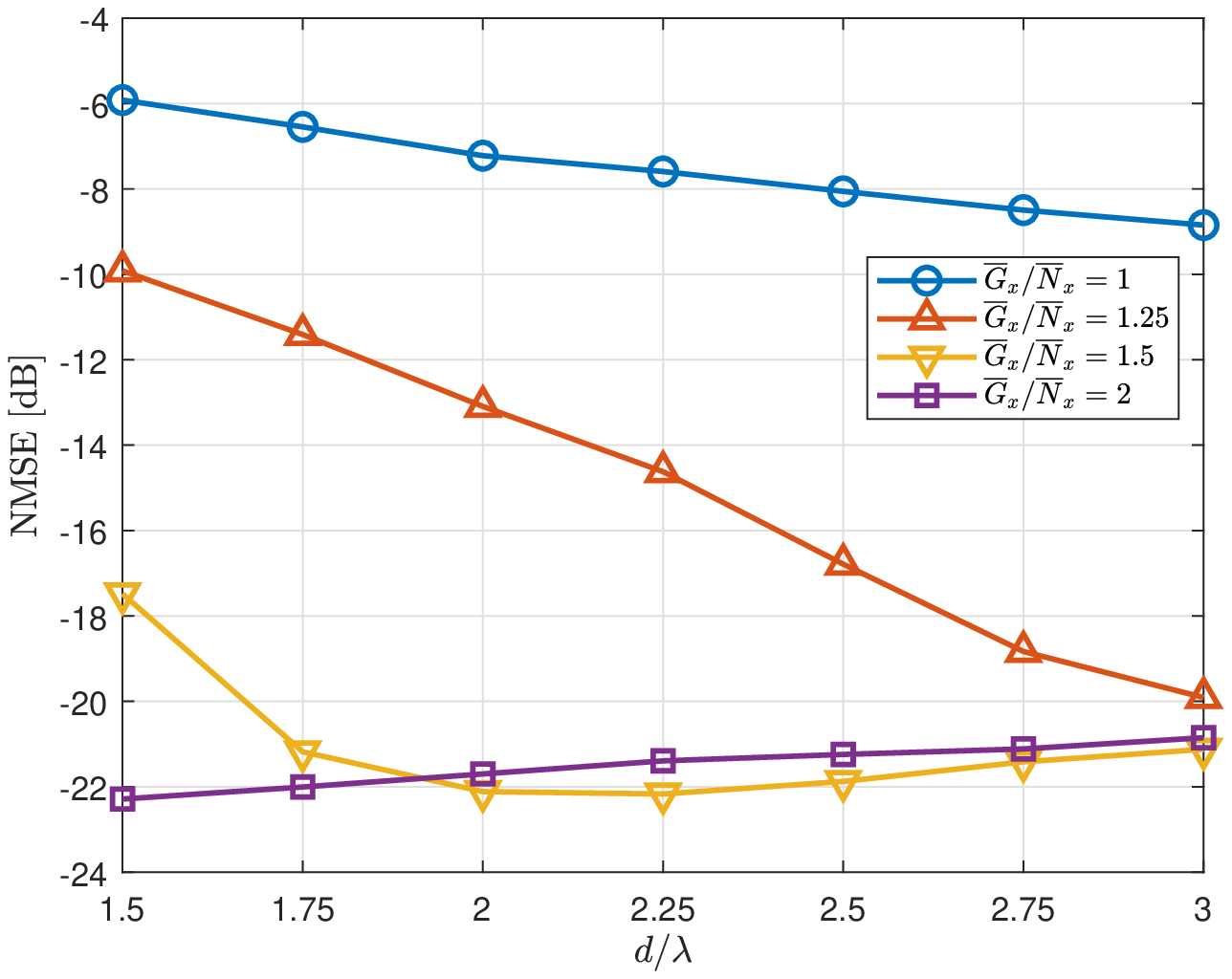}
\vspace*{-3mm}
\caption{The performance of the proposed scheme vs. the inter-element spacing of WSA. The number of iterations is $100$.}
\label{FigSpacing} 
\end{minipage}}
\vspace{-8mm}
\end{figure}

\begin{figure}[tp!]
\vspace*{-7mm}
\captionsetup{font={footnotesize}, name = {Fig.}, singlelinecheck=off, labelsep = period}
\begin{minipage}[t]{0.49\linewidth}
\centering
\includegraphics[width=2.6in]{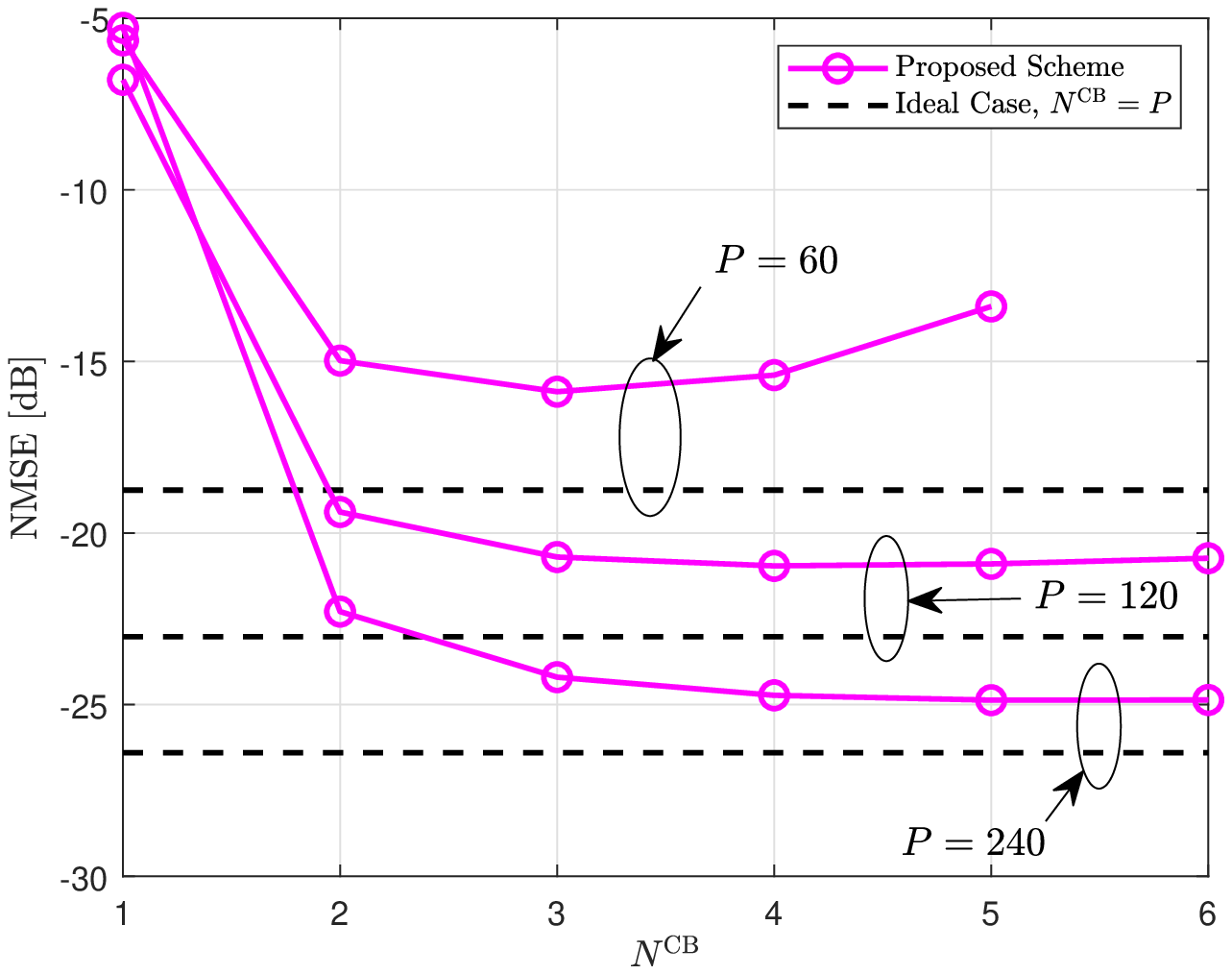}
\vspace*{-4mm}
\caption{The radar sensing performance with the proposed waveform design. The idealized case having no configurable time \cite{TcomSurvey} is considered as the lower bound.}
\label{FigNcb} 
\end{minipage}
\hfill
\begin{minipage}[t]{0.49\linewidth}
\centering
\includegraphics[width=2.6in]{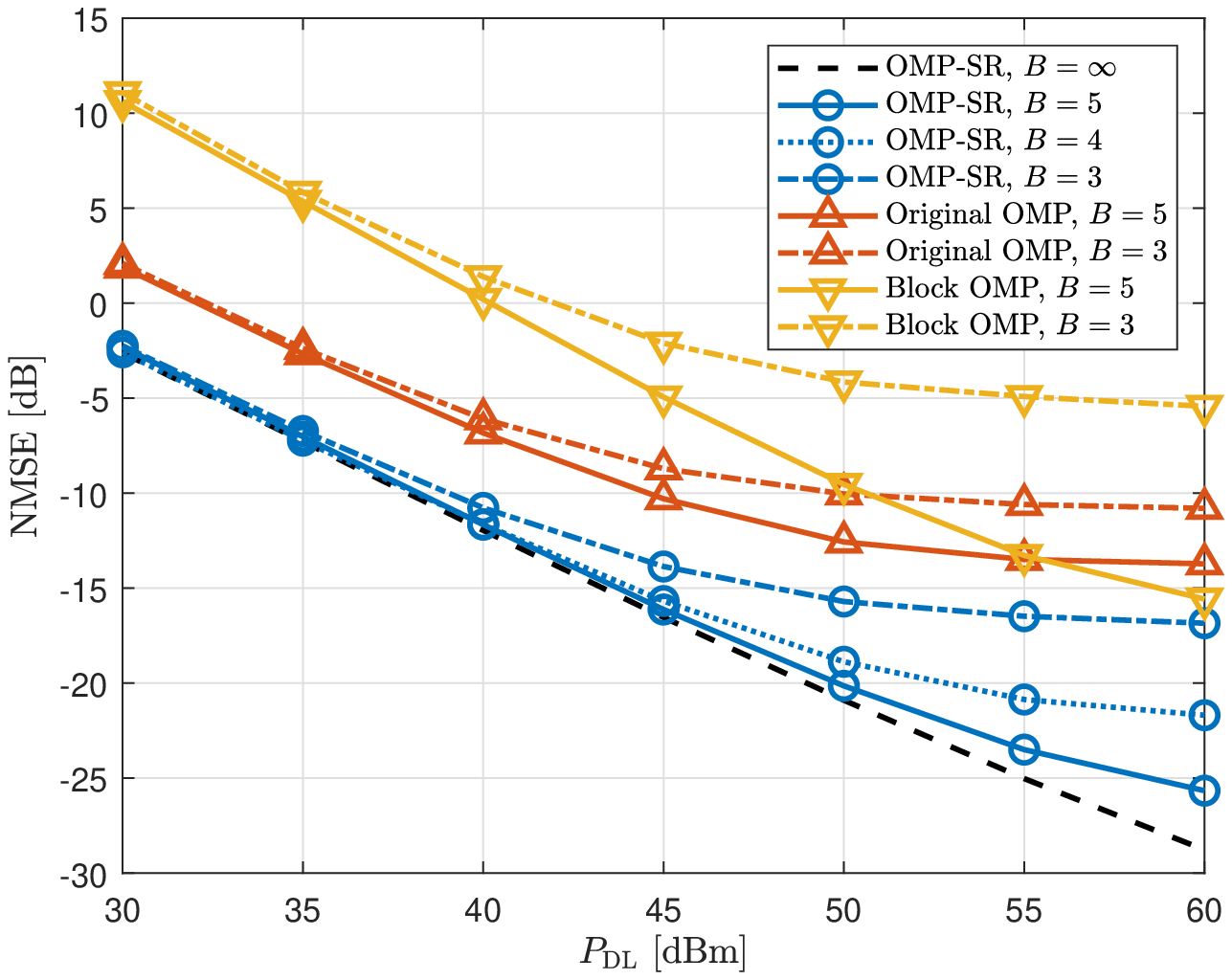}
\vspace*{-4mm}
\caption{Performance comparison of different radar CIR recovery algorithms, including the proposed OMP-SR, the original OMP \cite{korean}, and the block OMP \cite{XMa}, with $P=290$.}
\label{FigCompare} 
\end{minipage}
\vspace{-8mm}
\end{figure}

For radar sensing, the dimension of the dictionary for the CU is fixed to $G_x^{\rm CU} = N_x$. In Figs.~\ref{FigIter} to \ref{FigCompare}, we investigate the performance of the proposed radar sensing scheme by evaluating the normalized mean square error (NMSE) between the real radar CIR $\overline{\bf{H}}_{\rm SD}$ and its estimate $\widehat{\overline{\bf{H}}}_{\rm SD}$, which is given by $\textsf{E}\left\{ \frac{\left\| \widehat{\overline{\bf{H}}}_{\rm SD} - \overline{\bf{H}}_{\rm SD} \right\|_F^2}{\left\| \overline{\bf{H}}_{\rm SD} \right\|_F^2} \right\}$.

Specifically, Fig.~\ref{FigIter} depicts the convergence of the OMP-SR algorithm under different channel conditions. It can be seen that \textbf{Algorithm~\ref{ALG1}} converges reasonably fast and it achieves good NMSE performance under different channel conditions. In particular, Fig.~\ref{FigIter} reveals that initially the NMSE decreases rapidly as the iteration increases. After reaching the minimum NMSE value, further increase in the algorithm iteration degrades the NMSE performance, as too many iterations make the algorithm incapable of fitting the sparsity level of the actual CIR. {The results of Fig.~\ref{FigIter} also provide insight in choosing appropriate stop criterion for the OMP-SR algorithm. In particular, for scenarios where the CIR exhibits extreme sparsity, such as aerial target detection or satellite communications, the number of iterations should be small, while for terrestrial scenarios with more targets or scatters, the number of iterations should be moderately large.}

Fig.~\ref{FigSpacing} reports the NMSE performance as the function of inter-element spacing of the WSA $d$ under different dictionary dimensions $\overline{G}_x$, where $\overline{N}_x d > N_x d_{\rm cri}$, i.e., $d > \lambda$. Observe that within certain regime, e.g., $\overline{G}_x/\overline{N}_x < 1.5$, increasing $\overline{G}_x$ and $d$ significantly improves the NMSE performance of radar sensing, since larger $\overline{G}_x$ or $d$ results in finer angular resolution according to \eqref{WSASamples}. By contrast, for $\overline{G}_x/\overline{N}_x \ge 1.5$, the NMSE performance exhibits no improvement with increasing $d$. Note that increasing $\overline{G}_x$ enlarges the dimension of the CS problem \eqref{SparsePro}, while increasing $d$ would lead to bulky antenna array. Therefore, the values of $\overline{G}_x$ and $d$ should be carefully chosen to strike a balance between system performance and hardware complexity.

To investigate the proposed pilot waveform design, we plot the NMSE performance as the function of the codebook size $N^{\rm CB}$ in Fig.~\ref{FigNcb}. We also depict the idealized case where the analog precoders could change in each sample without reconfigurable time, i.e., $N^{\rm CB} = P$, as done in \cite{TcomSurvey}, which forms the lower bound of the NMSE\footnote{In practice, this idealized lower bound is unrealizable, see \textbf{Remark~\ref{REM1}.}}. As expected, the radar sensing suffers from the limited pilot diversity when $N^{\rm CB} = 1$, since the beam pattern produced by a single analog precoder during radar sensing is very likely to miss the real position of targets. Hence, it is necessary to increase $N^{\rm CB}$ to obtain higher pilot diversity for better sensing performance, as shown in Fig.~\ref{FigNcb}. For typical values of $P$, e.g., $120$ and $240$, the sensing performance improves with $N^{\rm CB}$ quickly, reaching the NMSE performance very close to the idealized lower bound. However, for small $P$, e.g., $60$, increasing $N^{\rm CB}$ may cause performance loss. This is because larger $N^{\rm CB}$ would require more zero pilot signals (see Fig.~\ref{FigWFdesign}) and thus degrade the received power. In practice, we can choose some appropriate value of $N^{\rm CB}$, e.g., $3$, for not only achieving good sensing performance but also alleviating the storage burden at the DFRC station.

Fig.~\ref{FigCompare} compares the performance of the proposed OMP-SR algorithm with two existing schemes, the original OMP algorithm \cite{korean} without the support refinement, and the block-OMP algorithm \cite{XMa} which only utilizes the delay-domain sparsity. The number of iterations for the original OMP is set to that of the OMP-SR for fare comparison, while the number of iterations for the block OMP is fixed to $10$, as it only considers the delay sparsity but not angular sparsity. The impact of low-resolution ADCs is also investigated in Fig.~\ref{FigCompare}. It can be seen that the OMP-SR outperforms the other two schemes significantly in the whole range of downlink transmit power and for different quantization bits, as it takes full advantage of the higher angular resolution of the WSA and the spatial consistency. Moreover, Fig.~\ref{FigCompare} indicates that the performance loss caused by low-resolution ADC is acceptable for our OMP-SR algorithm. For example, the sensing performance with practical $5$-bit ADCs can well approach that with the ideal infinite-resolution ADCs (the black dotted line), and this further validates the effectiveness of our proposed radar receiver. Also note that $P=290$ in Fig.~\ref{FigCompare} makes the dimension of measurements $Q\overline{N}=2568$ much smaller than that of actual CIR $LN\overline{N}=4096$. Consequently, the matrix ${\bf \overline{\Phi}}^{\rm H}{\bf \overline{\Phi}}$ is not invertible, and thus the conventional linear estimation methods, such as least squares, fail to work properly. This reveals the necessity of CS in the face of under-determined measurements.

\begin{figure}[tp!]
\vspace{-1mm}
\centering
\captionsetup{font={footnotesize}, name = {Fig.}, singlelinecheck=off, labelsep = period}
\subfigure[Performance with OMP-SR.]{\includegraphics[width=2.6in]{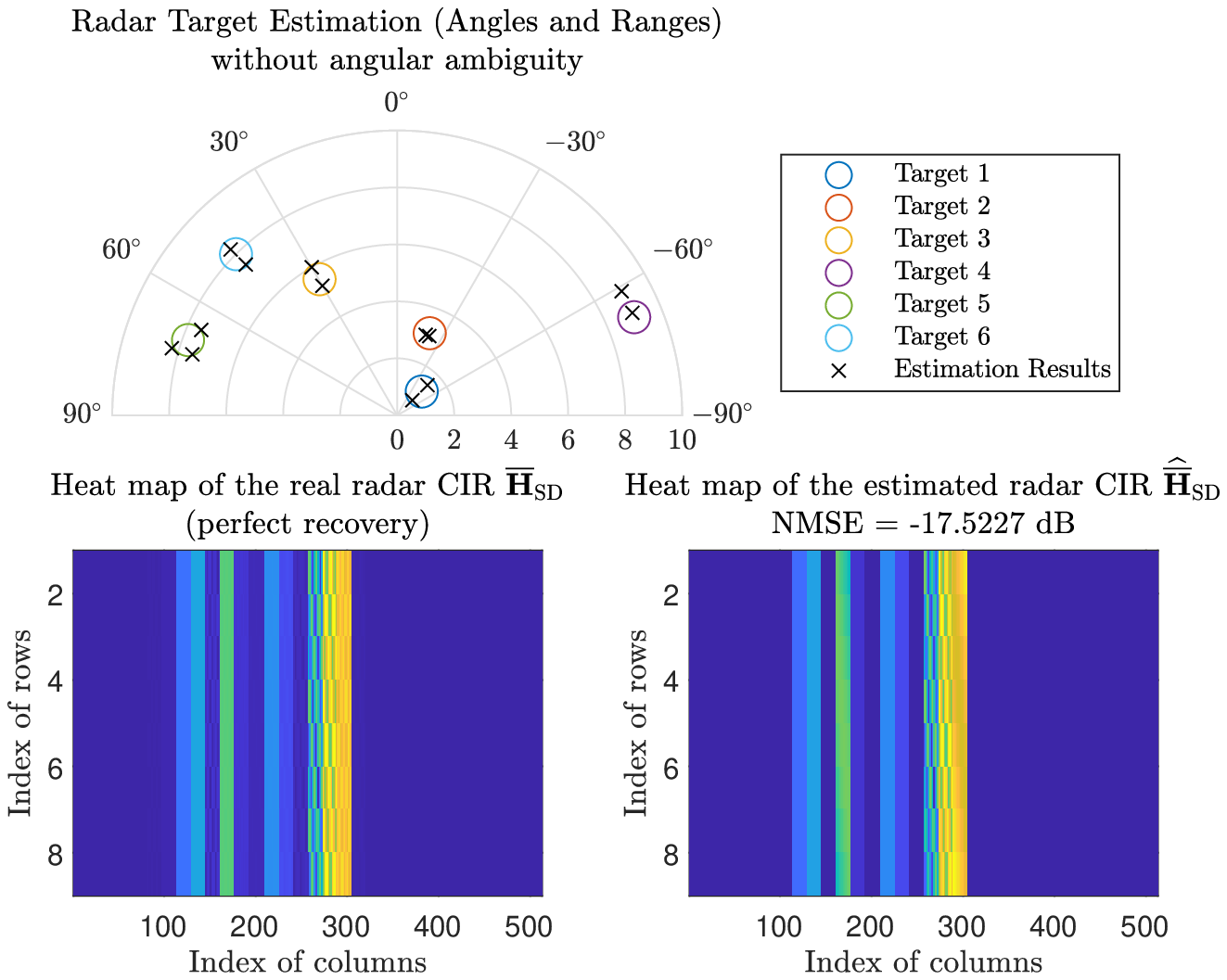}}
\hspace{5mm}
\subfigure[Performance without support refinement.]{\includegraphics[width=2.6in]{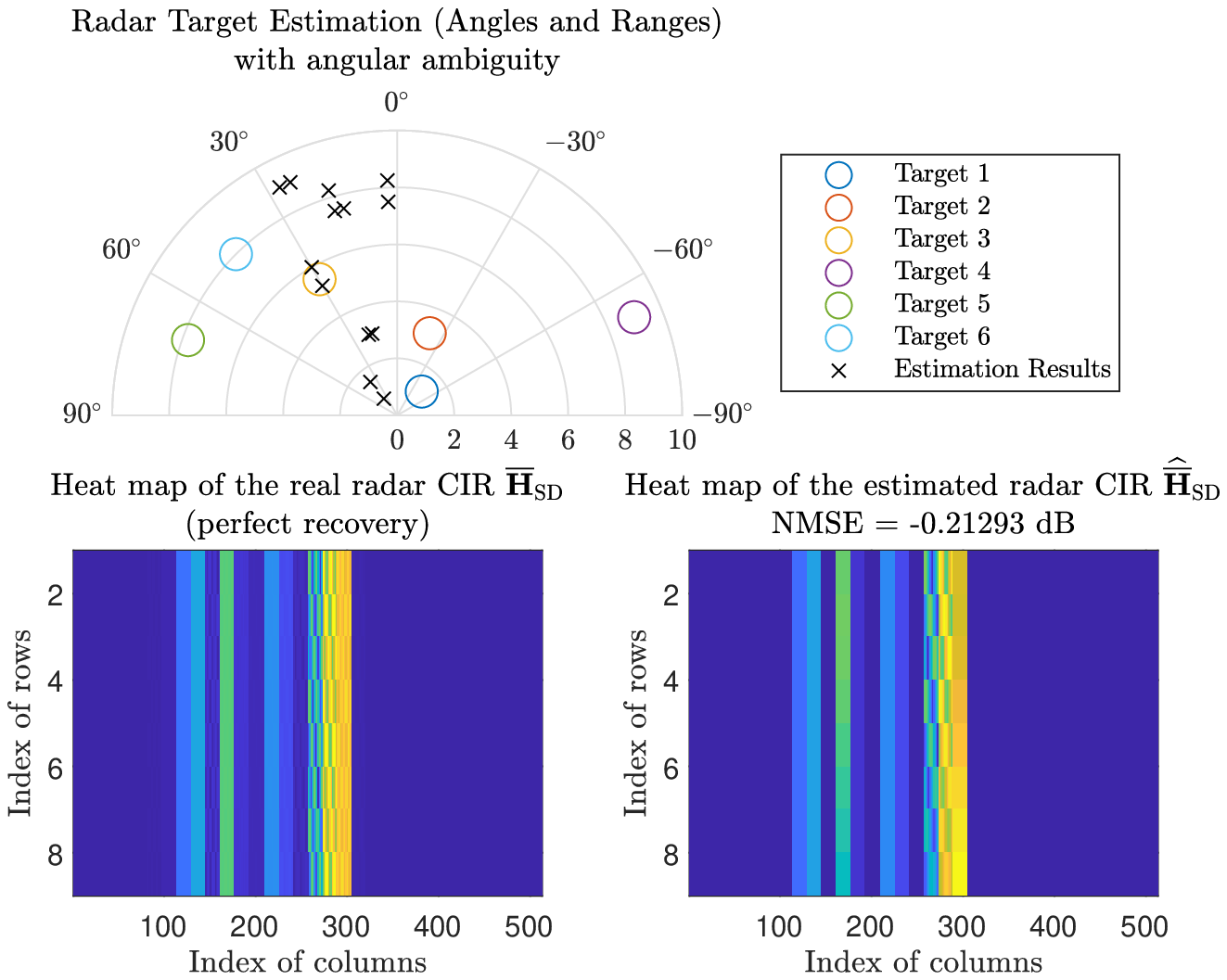}}
\vspace{-5mm}
\caption{A visualization of radar sensing performance. When plotting the estimated positions of targets, we only reserve the components whose amplitudes are larger than $\frac{\sigma_{\rm n}}{\sqrt{N_{\rm C}}}$. The CIRs in the spatial-delay domains are also presented.}
\label{FigVisual} 
\vspace{-8mm}
\end{figure}

To intuitively show the performance of radar sensing, we compare the estimated results to the true values of the angle and range (delay) parameters for a single channel realization in Fig.~\ref{FigVisual}. For clearness purpose, we consider point targets, i.e., $\overline{N}_{\rm P}\! =\! 1$, and set the number of iterations to $20$ for \textbf{Algorithm~\ref{ALG1}}. It can be seen from Fig.~\ref{FigVisual}\,(a) that by using our OMP-SR algorithm, all the targets' angles and delays (ranges) can be accurately estimated. By contrast, if we only use the angle estimation from step 10 of \textbf{Algorithm~\ref{ALG1}} without support refinement, i.e., use the original OMP, although the delay estimation is as accurate as that obtained by OMP-SR algorithm, the angle estimation exhibits severe blurring due to the angular ambiguity induced by the WSA, as can be seen clearly from Fig.~\ref{FigVisual}\,(b). This blurring will result in missing detection and/or false alarm in radar sensing, putting the served UTs at risk of collision and/or sudden stop. Therefore, the proposed support refinement procedure is necessary when WSA is considered.

\subsubsection{Communication performance}\label{S5.2.2}

Next, we investigate the communication performance under the proposed ISAC framework. We adopt the average spectral efficiency (ASE) to evaluate the LoS angle estimation performance of the low-complexity \textbf{Algorithm~\ref{ALG2}}. The ASE is defined by
\begin{align}
\text{ASE} =& \textsf{E} \left\{ \frac{1}{N_{\rm D}} \sum\limits_{n = 1}^{N_{\rm{D}}} \log_2\left( 1 + \frac{P_{\rm{DL}}}{\sigma_{\rm{n}}^2 N_{\rm{RF}}} \left| \left[ \widehat{\bf{a}}_{\rm{UT}}^{\rm H} {\bf{H}}_{\rm{SF}} \left( {\bf{I}}_{N_{\rm{D}}} \otimes \widehat{\bf{a}}_{\rm{CU}} \right) \right]_n \right|^2 \right)  \right\} \, [{\rm{bit/s/Hz}}] \, ,
\end{align}
where $\widehat{\bf{a}}_{\rm{UT}} = {\bf{a}}\left( \widehat{\mu}_{\rm{UT}};M_x \right)$ and $\widehat{\bf{a}}_{\rm{CU}} = {\bf{a}}\left( \widehat{\mu}_{\rm{CU}};N_x \right)$ are the steering vectors towards the estimated LoS directions $\widehat{\mu}_{\rm{UT}}$ and $\widehat{\mu}_{\rm{CU}}$, respectively, ${\bf{H}}_{\rm{SF}} = \left[ {\bf{H}}_{\rm{SD}} ~ {\bf{0}}_{{M_x} \times \left( N_{\rm{D}} - L \right) N_x} \right] \left( {\bf{F}}_{\rm{DFT}} \otimes {\bf{I}}_{N_x} \right)$ is the spatial-frequency (SF)-domain channel, and ${\bf{F}}_{\rm{DFT}}$ is the $N_{\rm D} \times N_{\rm D}$ DFT matrix. We also define $r_{\rm dic} {\buildrel \Delta \over =}  G_x^{\rm CU}/N_x = G_x^{\rm UT}/M_x$ as a dictionary design parameter. 
 
\begin{figure}[t]
\vspace*{-2mm}
\captionsetup{font={footnotesize}, name = {Fig.}, singlelinecheck=off, labelsep = period}
{\begin{minipage}[t]{0.49\linewidth}
\centering
\includegraphics[width=2.6in]{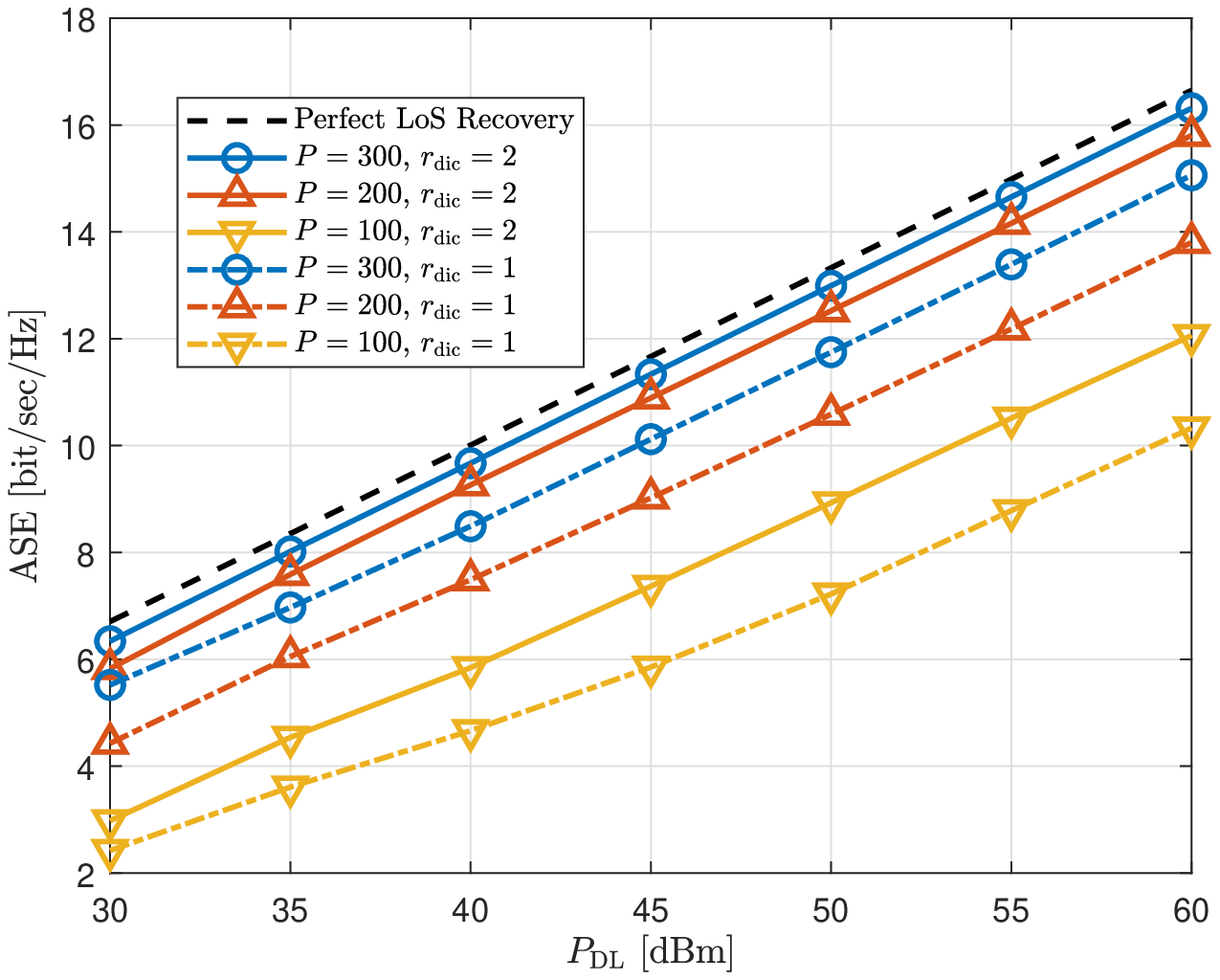}
\vspace*{-4mm}
\caption{ASE performance versus downlink transmit power.}
\label{FigASE} 
\end{minipage}}
\hfill
{\begin{minipage}[t]{0.49\linewidth}
\centering
\includegraphics[width=2.6in]{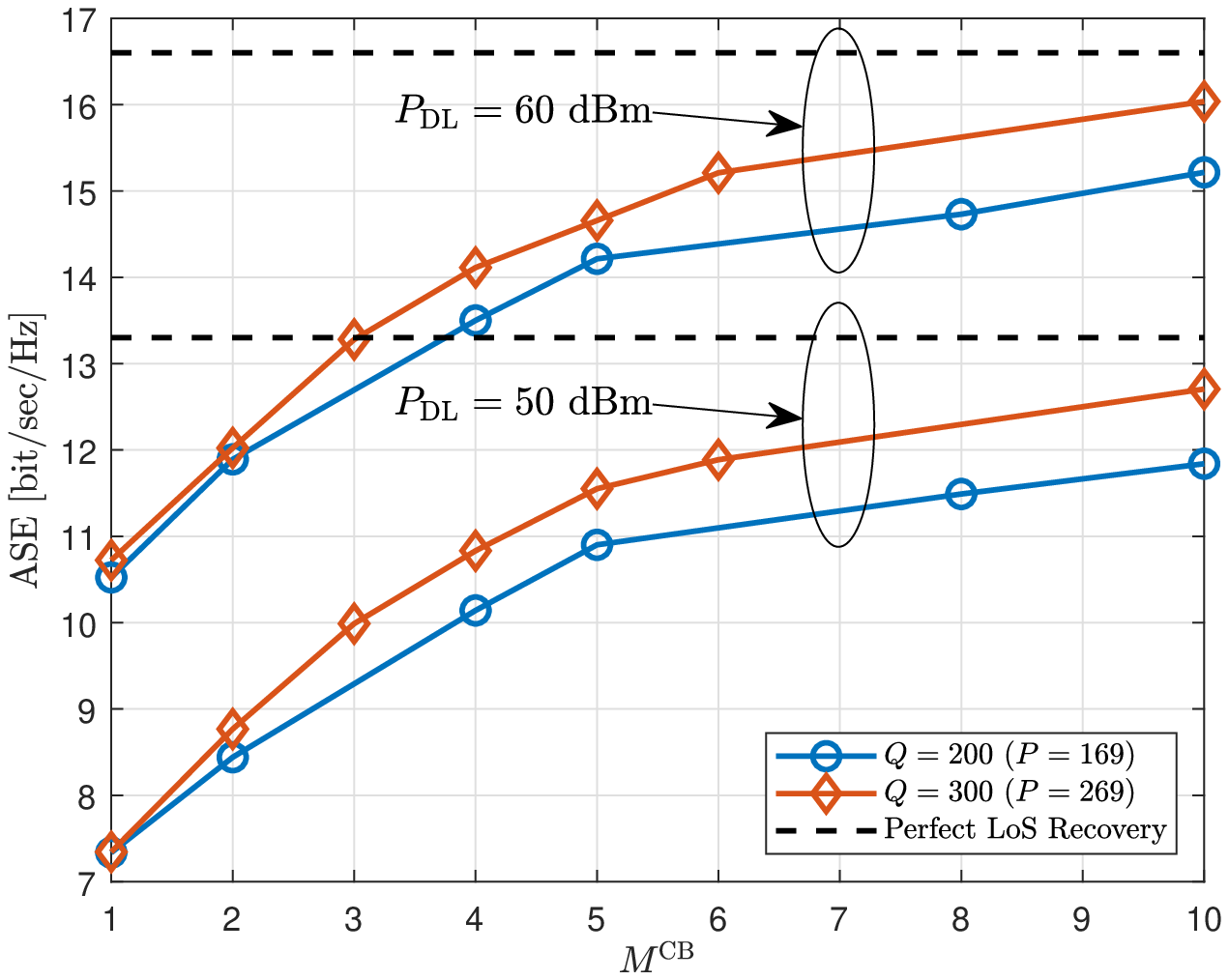}
\vspace*{-4mm}
\caption{ASE performance versus UT codebook size $M^{\rm CB}$.}
\label{FigMcb} 
\end{minipage}}
\vspace{-8mm}
\end{figure}

To validate the effectiveness of the proposed low-complexity CE scheme, Fig.~\ref{FigASE} depicts the ASE performance, based on the estimated LoS angles, as a function of the downlink transmit power, against various values of $r_{\rm dic}$. It can be seen that compared with the non-redundant dictionary ($r_{\rm dic}\! =\! 1$), our redundant dictionary design ($d_{\rm dic}\! >\! 1$) significantly improves the ASE performance given the same pilot overhead. With $P=300$ and $r_{\rm dic}\! =\! 2$, the ASE performance well approaches that with the perfect LoS angles. Due to its superior performance, we adopt $r_{\rm dic} = 2$ to investigate the impact of our pilot waveform design. Fig.~\ref{FigMcb} plots the ASE performance as the function of UT codebook size $M^{\rm CB}$. For simplicity, we fix $Q$ for each curve in Fig.~\ref{FigMcb} and vary $T_{\rm RF}^{\rm UT}$ to obtain different $M^{\rm CB}$. Similar to Fig.~\ref{FigNcb}, Fig.~\ref{FigMcb} shows that a single analog combiner ($M^{\rm CB}\! =\! 1$) leads to poor ASE performances, while increasing $M^{\rm CB}$ significantly improves the ASE performances. The results of Figs.~\ref{FigNcb} and \ref{FigMcb} confirm that our pilot waveform design is effective and necessary for HBF-aided ISAC systems to realize pilot diversity under practical hardware constraint.

\begin{figure}[t]
\vspace*{-2mm}
\captionsetup{font={footnotesize}, name = {Fig.}, singlelinecheck=off, labelsep = period}
\begin{minipage}[t]{0.49\linewidth}
\centering
\includegraphics[width=2.6in]{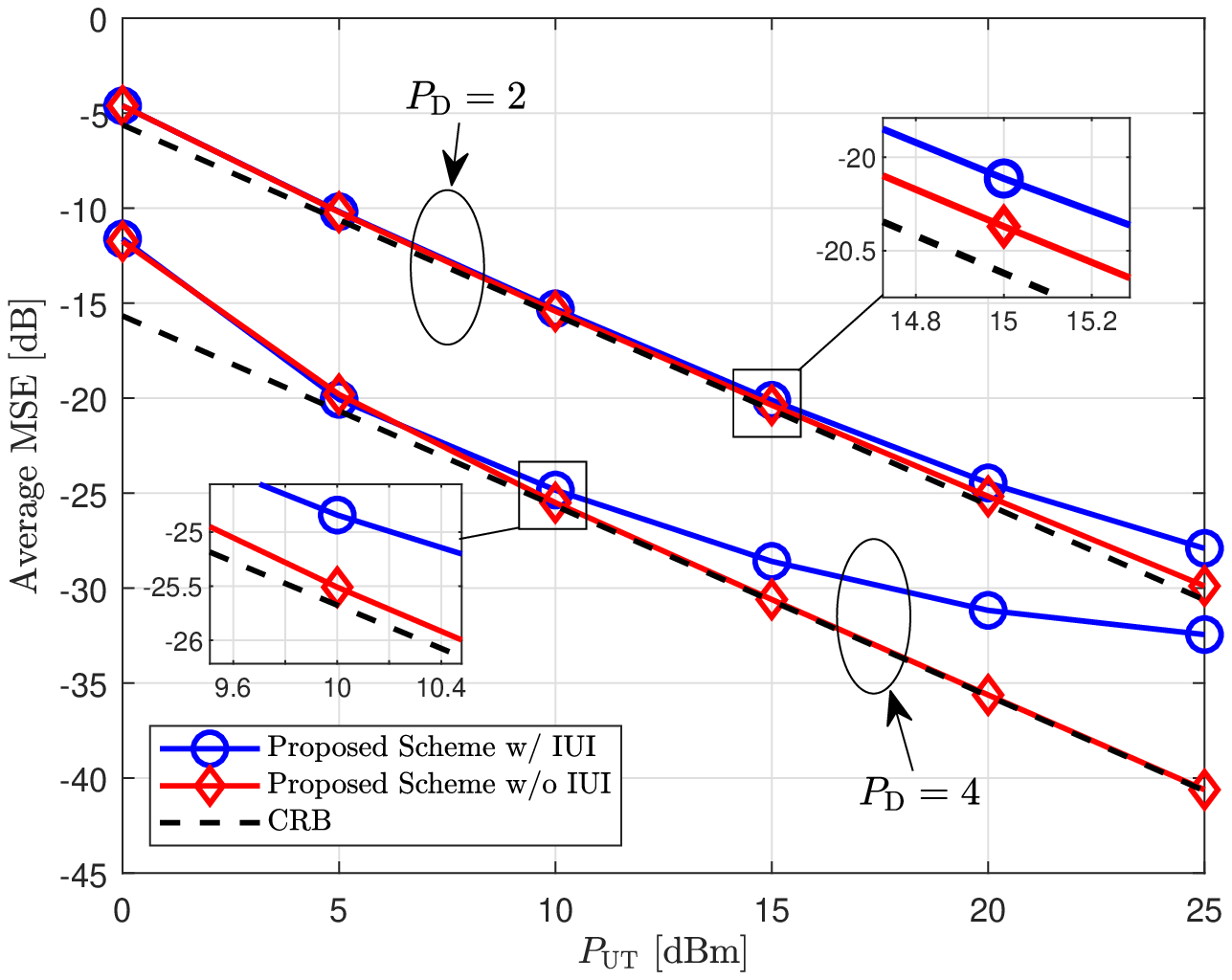}
\vspace*{-4mm}
\caption{Doppler estimation performance. For the initial CE, $P = 300$. For multi-UT Doppler estimation, $U = 4$, and the transmit power of each UT is equal (denoted as $P_{\rm UT}$).}
\label{FigDoppler} 
\end{minipage}
\hfill
\begin{minipage}[t]{0.49\linewidth}
\centering
\includegraphics[width=2.6in]{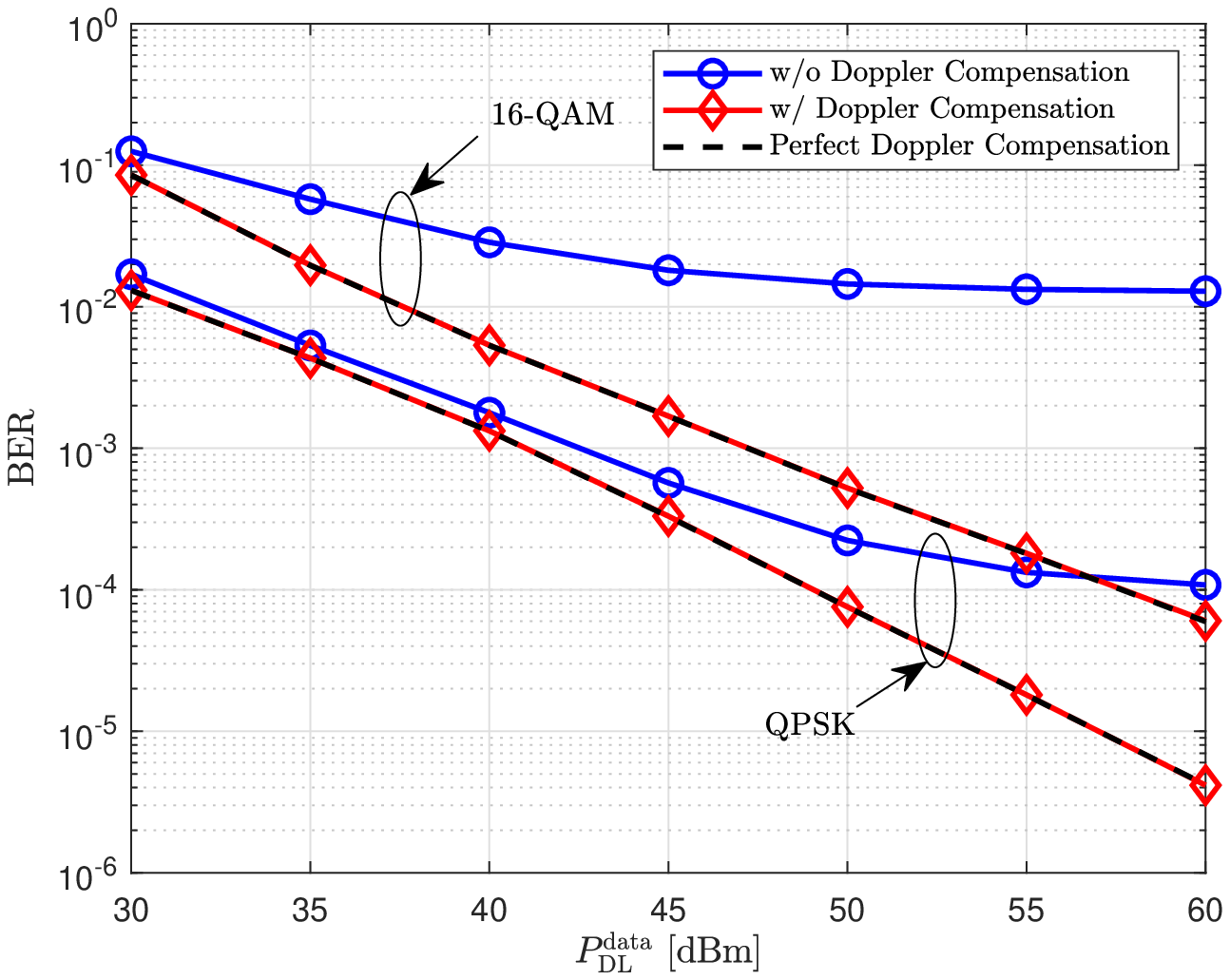}
\vspace*{-4mm}
\caption{Downlink BER performance. The same scenario in Fig.~\ref{FigDoppler} is considered. The estimation results with $P_{\rm UT} = 23$ dBm and $P_{\rm D} = 4$ are adopted for Doppler compensation.}
\label{FigBER} 
\end{minipage}
\vspace{-8mm}
\end{figure}

We further investigate the effectiveness of the proposed Doppler estimation scheme. Note that we only reserve the results for those UTs with $\left\| {{\bf{y}}_u^{\rm{D}}} \right\|_2^{} \ge 10\sqrt {{P_{\rm{D}}}} {\sigma _{\rm{n}}}$\footnote{This is because the very low energy of ${{\bf{y}}_u^{\rm{D}}}$ in \eqref{TS} may indicate the unreliable CE results at the initial estimation stage. In such a case, re-estimation of angles and delays is necessary before conducting Doppler estimation.}. Fig.~\ref{FigDoppler} compares the average MSE performance of Doppler estimation with or without IUI. The average MSE is defined as
$\textsf{E}\left\{ \sum\limits_{u = 1}^U \left| 2\pi T_{\rm{D}}\left( \widehat{f}_{{\rm{D}},u} - f_{{\rm{D}},u} \right) \right|^2 \right\}$ \cite{WNALP,CRLB}, where $f_{{\rm{D}},u}$ is the Doppler frequency of the $u$-th UT and $\widehat{f}_{{\rm{D}},u}$ is its estimate. We also plot the CRB of the single-tone frequency estimation problem \cite[(12)]{CRLB} in Fig.~\ref{FigDoppler}. It can be seen that for $P_{\rm D} = 2$, the influence of IUI is not serious and it hardly affects the Doppler estimation. Hence, the MSE performance with IUI can attain the CRB, particularly in the low transmit power regime of $P_{\rm UT} < 10$\,dBm. By contrast, for larger $P_{\rm D}$, e.g., $P_{\rm D} = 4$, the MSE performance deviates from the CRB as $P_{\rm UT}$ increases, leading to an error floor. However, this MSE gap only makes negligible influence on the real communication performance as will be seen next.

Fig.~\ref{FigBER} plots the downlink bit-error-rate (BER) performance with or without Doppler compensation. We consider the orthogonal frequency division multiplexing (OFDM) transmission with $N_{\rm D}\! =\! 1024$ sub-carriers between the CU and UTs. The estimated Doppler frequency obtained by the proposed scheme with $P_{\rm UT}\! =\! 23$\,dBm and $P_{\rm D}\! =\! 4$ is used to compensate for the Doppler effect of the LoS path. It can be seen from Fig.~\ref{FigBER} that the BER with the estimated Doppler compensation is almost the same as that with the perfect Doppler compensation, even though the MSE performance cannot achieve the CRB, as shown in Fig.~\ref{FigDoppler}. Without Doppler compensation, severe inter-carrier interference caused by Doppler effect \cite{Book} degrades the BER dramatically. Specifically, it can be seen from Fig.~\ref{FigBER} that with Doppler compensation, there is about $10$\,dB gain at the BER of $10^{-4}$ for uncoded $16$-QAM modulation scheme. This means that the proposed scheme for Doppler estimation and compensation is vital for combating time-varying mmWave channels in communication-centric ISAC systems.

\section{Conclusions}\label{S6}

We have investigated the ISAC system aided by mmWave mMIMO with HBF architecture. First, we have introduced an energy-efficient WSA architecture as the radar receiver to enhance the angular resolution of radar sensing. Then, we have designed an ISAC frame structure for time-varying ISAC systems, which facilitates the estimation of angles, delays, and the Doppler frequencies. In particular, the pilot waveforms have been designed to meet the hardware constraints induced by HBF array. In order to reduce the pilot overhead, we formulated the ISAC processing as sparse signal recovery problems with dedicated dictionaries, to utilize advanced compressive sensing techniques. Specifically, we have proposed the orthogonal matching pursuit with support refinement algorithm, which can cope with angular ambiguity and achieve better recovery performance than its traditional counterparts. We also provided a framework of estimating the Doppler frequencies of users/targets, which is essential for both speed measurement and payload data demodulation. Possible future research directions based on this paper include the study of robust quantized CS methods, the interaction between the radar sensing algorithm and CE algorithm, the beamforming design for the proposed transceiver architecture, the analysis of near-field effect, and the proof-of-concept field experiments.

\end{document}